\documentclass[twocolumn,usenatbib]{mnras}

\usepackage{lmodern}
\usepackage{multicol}
\usepackage{graphicx}
\usepackage{bookmark}
\usepackage{tabularx}
\usepackage[T1]{fontenc}
\usepackage{ae,aecompl}
\usepackage{pdflscape}
\usepackage{subfig}
\usepackage{float}

\usepackage{savesym}
\savesymbol{iint}
\usepackage{graphicx}	
\usepackage{amsmath}	
\usepackage{amssymb}	
\usepackage{etoolbox}

\title[]{UOCS\thanks{UVIT Open Cluster Study} -- VII. Blue Straggler Populations of Open Cluster NGC 7789 with UVIT/AstroSat}

\author[K. Vaidya et al.]{
Kaushar Vaidya$^{1}$,\thanks{E-mail: kaushar@pilani.bits-pilani.ac.in}
Anju Panthi$^{1}$,
Manan Agarwal$^{1}$,
Sindhu Pandey$^{2}$,
Khushboo K. Rao$^{1}$,
\newauthor
Vikrant Jadhav$^{3,4}$,
Annapurni Subramaniam$^{3}$
\
\\
$^{1}$Department of Physics, Birla Institute of Technology and Science - Pilani, 333031, Rajasthan, India\\
$^{2}$Aryabhatta Research Institute of Observational Sciences, Manora Peak, Nainital, India \\
$^{3}$Indian Institute of Astrophysics, Sarjapur Road, Koramangala, Bangalore, India \\
$^{4}$Joint Astronomy Programme and Department of Physics, Indian Institute of Science, Bangalore, India \\
}

\date{Accepted XXX. Received YYY; in original form ZZZ}

\pubyear{2021}

\begin{document}
\label{firstpage}
\pagerange{\pageref{firstpage}-\pageref{lastpage}}
\maketitle

\begin{abstract}
NGC 7789 is a $\sim$1.6 Gyr old, populous open cluster located at $\sim$2000 pc. We characterize the blue straggler stars (BSS) of this cluster using the Ultraviolet (UV) data from the UVIT/$AstroSat$. We present spectral energy distributions (SED) of 15 BSS candidates constructed using multi-wavelength data ranging from UV to IR wavelengths. In 8 BSS candidates, a single temperature SED is found to be satisfactory. We discover hot companions in 5 BSS candidates. The hot companions with T$\mathrm{_{eff}} \sim$11750--15500 K, $R$ $\sim$0.069--0.242 R$\mathrm{_{\odot}}$, and $L$ $\sim$0.25--1.55 L$\mathrm{_{\odot}}$, are most likely extremely low mass (ELM) white dwarfs (WDs) with masses smaller than $\sim$0.18 M$\mathrm{_{\odot}}$, and thereby confirmed post mass transfer systems. We discuss the implication of this finding in the context of BSS formation mechanisms. Two additional BSS show excess in one or more UV filters, and may have a hot companion, however we are unable to characterize them. We suggest that at least 5 of the 15 BSS candidates (33\%) studied in this cluster have formed via the mass-transfer mechanism.
\end{abstract}




\section{Introduction} \label{Section 1}
Blue straggler stars (BSS) are one of the exotic stellar populations whose evolution differs from normal single stars.  They are commonly found in diverse stellar environments such as globular clusters \citep{sandage53, fusi92, sarajedini93}, open clusters \citep{johnson55,burbidge58,sandage62,rain21}, dwarf galaxies \citep{momany07,mapelli09}, and galactic fields \citep{preston00}.  There are three fundamental mechanisms that explain the formation of BSS: (a) direct stellar collisions in dynamical interactions in dense stellar environments \citep{hills76}, (b) mass transfer in close binary systems \citep{mccrea64}, and (c) merger or mass transfer of an inner binary in a triple system through Kozai mechanism \citep{perets09}. The stellar collisions of binaries with single stars or another binary in dynamical interactions may result into single or binary BSS \citep{leonard89, ferraro95, leigh11}. Mass transfer that occurs when the primary is on the main-sequence may result into a single massive BSS or a binary BSS with a short-period main-sequence companion \citep[Case A;][]{webbink76}. Roche lobe overflow mass transfer in a binary system that occurs when the primary star is on the
red-giant branch, i.e. RGB, will likely produce a short-period binary with a Helium (He) WD of mass $\leq$ 0.45 M$\mathrm{_{\odot}}$ as a companion \citep[Case B;][]{mccrea64}, whereas when the primary star is on the asymptotic-giant branch, i.e. AGB, will result into a long-period binary with a CO WD of mass $\geq$ 0.5 M$\mathrm{_{\odot}}$ as a companion \citep[Case C;][]{chen08,gosnell14}. A comprehensive overview of BSS in diverse environments and their formation channels can be found in this review by \citet{boffin15}.

The relative importance of various BSS formation mechanisms in diverse stellar environments continues to be a significant topic of interest. UV wavelengths offer a specific advantage in such an investigation by enabling the detection of BSS binaries with hot companions on the basis of the excess emission in the UV wavelengths. \citet{gosnell15} detected WD companions in 7 out of 15 single lined spectroscopic binaries (SB1) of the open cluster NGC 188 using the far-UV (FUV) data from the Hubble Space Telescope (HST). They inferred that these 7 BSS were formed through recent mass transfer.  Using the Ultraviolet Imaging Telescope (UVIT) onboard the $AstroSat$, WD companions to BSS of several star clusters have been identified including NGC 188 \citep{subramaniam16}, M67 \citep{sindhu19, jadhav19, sindhu20, pandey21}, and the globular cluster NGC 5466 \citep{sahu19}. In the globular cluster NGC 1851, \citet{singh20} discovered extreme Horizontal Branch (EHB) stars as hot companions to the BSS using the $AstroSat$/UVIT system. This method of analysis allows constraining of the mass transfer and binary evolution models. 

NGC 7789 ($\alpha$ = 23$^h$57$^m$21$^s$.6, $\delta$ = +56$^\circ$43$\arcmin$22$\arcsec$ J2000) is a populous, 1.6 Gyr old open cluster with an estimated distance modulus $(m-M)_V \leq 12.2 $, and reddening $0.35 \leq E(V-I) \leq0.38$ \citep{gim98}. \citet{burbidge58} presented the oldest photometric study of the cluster. \citet{mcnamara81} identified 679 probable members brighter than B $\sim$15.5 on the basis of proper motion membership analysis. The earliest radial velocity measurements of giants were reported by \citet{thogersen93} and \citet{friel95}. Sixteen BSS candidates of the cluster were identified as possible members of the cluster based on the then existing proper motion, radial velocity, and polarization data by \citet{gim98a}.
By spectroscopic studies of the abundances of giant stars of the cluster, \citet{jacobson11} reported the metallicity, [Fe/H] = +0.02$\pm$0.04, and \citet{overbeek15} reported [Fe/H] = +0.03$\pm$0.07.
\citet{cantat18} estimated the distance of the cluster to be 2075 pc based on the Gaia DR2 membership analysis. Using the cluster members identified by \citet{cantat18} and \citet{cantat20}, a BSS catalogue of 408 open clusters was presented by \citet{rain21}. They reported 16 BSS candidates in NGC 7789. Recently, \citet{overlar20} found an extended main-sequence turnoff in several open clusters including NGC 7789 in their Gaia DR2 study. They reported the cluster age to be 1070$\pm$390 Myr. \citet{nine20} presented a time series radial velocity survey of this cluster obtained with Hydra Multi-Object spectrograph on the WIYN 3.5 m telescope in which they studied various stellar populations of the cluster including the main-sequence stars down to one magnitude below the turnoff. In particular, they identified 12 BSS as cluster members, of which, four (33\%) are found to be SB1.

NGC 7789 was observed by UVIT/$AstroSat$ under $AstroSat$ proposal A03-008. We study the BSS populations of NGC 7789 using the UV data from the $AstroSat$ and other archival data ranging from UV to IR wavelengths to characterize the BSS as well as to discover any hot companions based on the excess in their UV fluxes.
We present in Section \ref{Sec. 2}, the UVIT data and  analysis, in Section \ref{Section 3}, the results, in Section \ref{Section 4}, the discussions, and in Section \ref{Section 5}, the summary of the paper.

\section{Observations and Data Analysis} \label{Sec. 2}

The UVIT is one of the payloads on India's first multiwavelength space observatory $AstroSat$, that was launched on 28 September 2015. The UVIT has
two 38 cm telescopes: one has a single FUV channel that observes in 130 -- 180 nm wavelength range, the other has two channels, a near-UV (NUV) channel that observes in the 200 -- 300 nm wavelength range, and a visible (VIS) channel that observes simultaneously in the 350 -- 550 nm wavelength range. In each of the three channels, multiple sets of filters are available. The VIS channel is used for obtaining satellite drift corrections \citep{tandon17}. The UVIT generates simultaneous images in the FUV, NUV and VIS channels with a circular field of view of 28$\arcmin$ diameter and an angular resolution of < 1.8$\arcsec$ \citep{tandon17b}. The UV detectors operate in the photon counting mode, whereas the VIS detector operates in the integration mode . 

The UVIT data of NGC 7789 were obtained in July 2017 in a single FUV filter F169M, and in three NUV filters N245M, N263M, and N279N. We downloaded level-1 data of the cluster from the $AstroSat$ archive\footnote{https://astrobrowse.issdc.gov.in/astro\_archive/archive/Home.jsp}. We produced science-ready images by performing data reduction on raw images using CCDLAB \citep{postma17,postma20}. After that, we performed point-spread function (PSF) photometry on the images using DAOPHOT/IRAF tasks \citep{stetson87}. In order to obtain the UVIT magnitudes in the AB magnitude system, we used the zero-point (ZP) magnitudes reported in \citet{tandon20}. The PSF magnitudes were corrected for aperture corrections as well as for saturation correction as provided in \citet{tandon20}. Table \ref{table 1} gives information regarding the exposure times of the observations, the PSF, and the number of detections in all the UVIT filters.

\section{Results} \label{Section 3}

\subsection{The Color Magnitude Diagrams} \label{Section 3.1}
\citet{rao21} identified cluster members of NGC 7789 using Gaia DR2 proper motions and parallaxes \citep{brown18}. They identified 2799 cluster members within the estimated cluster radius 28$\arcmin$. \citet{nine20} presented a time-series radial velocity survey of 1206 sources within 18$\arcmin$ of the cluster center. They estimated cluster membership of these sources based on the radial velocity measurements of the sources obtained using Hydra/WIYN system and the proper-motion measurements from Gaia DR2. They identified a total of 624 cluster members including giants, red clump stars, BSS, sub-subgiants, and main-sequence stars down to one magnitude below the turnoff. 

\begin{figure*} 
\includegraphics[scale=0.6]{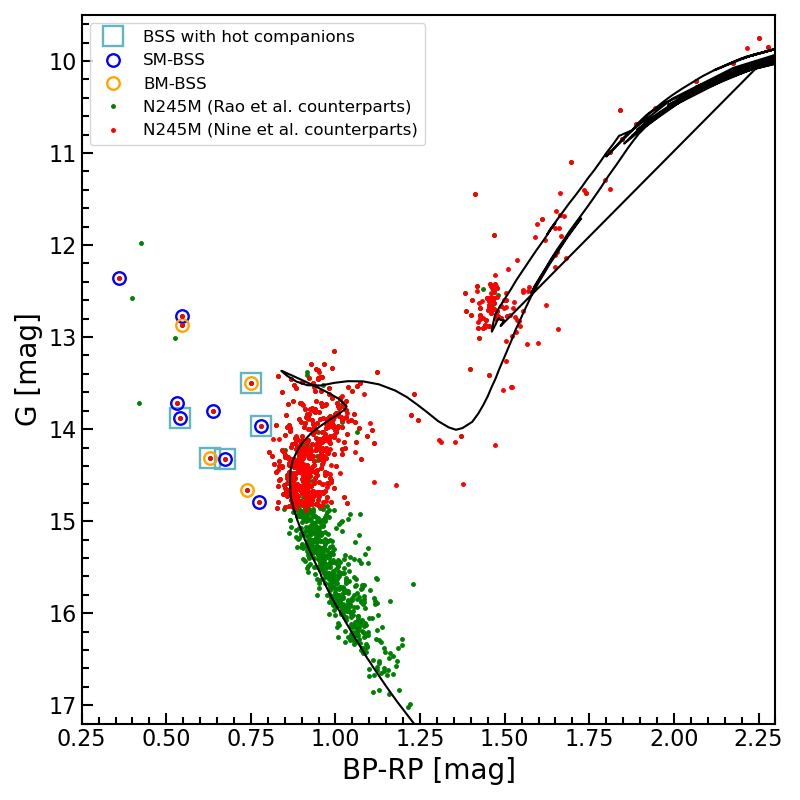} \par
\caption{Gaia DR2 CMD of N245M sources with a counterpart in the cluster members identified by \citet{rao21} as green dots, and \citet{nine20} as red dots. Known single and binary BSS from \citet{nine20} are shown as blue and orange open circles, respectively. BSS with hot companions are marked additionally as light blue open squares. A PARSEC isochrone of age = 1.6 Gyr, distance = 2000 pc, [Fe/H] = 0.023 has been plotted after applying the extinction correction of A${\mathrm{_G}}$ = 0.69, and reddening correction of E(${\mathrm{B_P-R_P}}$) = 0.375.}
\label{Fig. 1}
\end{figure*}

To characterize our UVIT detected sources further, we cross-matched them with the cluster members from \citet{rao21} and with the cluster members from \citet{nine20} by using a search radius of 1.5$\arcsec$ for the NUV filters and 2$\arcsec$ for the filter F169M. The information of the total number of members identified as counterparts in each UVIT filter is given in the Table \ref{table 1}. The Figure~\ref{Fig. 1} shows the Gaia DR2 color-magnitude diagram (CMD) of 1056 sources detected in the NUV filter N245M that are having a counterpart in the cluster members identified by \citet{rao21}. Among these, 497 sources are commonly identified as cluster members by \citet{nine20}. A PARSEC\footnote{http://stev.oapd.inaf.it/cgi-bin/cmd} isochrone \citep{bressan12} of age = 1.6 Gyr, distance = 2000 pc, [Fe/H]= 0.023 has been plotted after applying the extinction and reddening correction of A${\mathrm{_G}}$ = 0.69, and E(${\mathrm{B_P-R_P}}$) = 0.375. We detect all the 12 BSS members from \citet{nine20} in the NUV filters N245M and N263M, but do not detect 2 BSS in the NUV filter N279N and 3 BSS in the F169M filter. In addition, we detect 4 BSS candidates from \citet{rao21} in all the UVIT filters, that are marked on the Figure~\ref{Fig. 1}. The G-band magnitudes of the BSS range from $\sim$1 magnitude below the turnoff point to $\sim$2 magnitude above the turnoff point. Two BSS, a single member BSS that is also a rapid rotator, WOCS 25024, and a binary member BSS, WOCS 36011 \citep{nine20}, are seen to be located below the turnoff point, with somewhat bluer colors in the CMD. These two BSS may as well be termed as blue lurkers \citep[BL;][]{leiner19}, however, we refer to these two objects also as BSS instead of BL. We notice many RGB and red clump sources which are \citet{nine20} members, are having a detection in the N245M filter.    

\subsection{BSS Candidates} \label{Section 3.2}

\citet{nine20} identified 12 BSS candidates that are members according to both radial velocities and Gaia DR2 proper motions including 2 BSS with high noise proper motions. In addition, they listed 9 BSS candidates that are members according to either radial velocity ($P_{RV} \geq $ 50\%) or proper motions ($P_{\mu} \geq $ 50\%), but not according to both. All the 12 BSS that are members as per \citet{nine20}, are detected in two NUV filters N245M and N263M, whereas 10 of them are detected in the NUV filter N279N, and 9 are detected in the FUV filter F169M. In addition, we find 4 BSS candidates from \citet{rao21} in all the UVIT filters, three of which are found to be rapid rotators or very rapid rotators by \citet{nine20}. Upon comparing our 16 BSS candidates with the BSS candidates identified by \citet{rain21}, we find
9 BSS to be common. 4 BSS candidates identified by \citet{nine20} and 3 BSS candidates identified by \citet{rao21} are missing in \citet{rain21}. Most of their remaining 7 BSS candidates are located beyond 28$\arcmin$ from the cluster center, the radius of the cluster as per \citet{rao21}. These 7 BSS candidates are also beyond the area surveyed by \citet{nine20}, as well as the region surveyed in this work using the UV observations. Table~\ref{table 2} lists all the 16 BSS candidates with their UVIT and \textit{GALEX} NUV fluxes.

\subsection{Spectral Energy Distribution of BSS} \label{Section 3.3}
 
We construct the SED of the BSS candidates by using photometric data from UV to IR wavelengths. In order to construct the SEDs, we made use of the Virtual Observatory SED Analyzer \citep[VOSA;][]{bayo08}. We first compiled the photometric data of these sources in NUV from \textit{GALEX} \citep{martin05}, optical from \citet{gim98}, $Gaia$ EDR3 \citep{brown21}, and PAN-STARRS, near-IR from Two Micron All Sky Survey \citep[2MASS;][]{cohen03} and mid-IR from Wide-field Infrared Survey Explorer \citep[\textit{WISE};][]{wright10} using VOSA. Next, the fluxes were corrected for extinction according to the extinction law by \citet{fitzpatrick99} and \citet{indebetouw05} by providing the mean value of extinction $\mathrm{A_V}$ = 0.868 to VOSA. This mean extinction in the $V$ band was obtained from the mean value of $\mathrm{A_G}$ of the Gaia DR2 cluster members \citep{rao21} using the correlation available between the extinctions from \citet{wang19}. VOSA calculates synthetic photometry for selected filters for the chosen theoretical models which are then compared with the extinction corrected values of the fluxes of the sources using a chi-square test. The fitting process involves minimization of the value of reduced chi-square, $\chi^2_{r}$, which is determined using the following formula:
\begin{equation}
\chi^2_{r} = \frac{1}{N-N_f} \sum_{i=1}^{N} \frac{(F_{o,i}-M_dF_{m,i})^2}{\sigma_{o,i}^2}
\end{equation}
where $N$ is the number of photometric data points, $N_f$ is the number of free parameters in the model, $F_{o,i}$ is the observed flux, $M_dF_{m,i}$ is the model flux of the star, $M_d=(\frac{R}{D})^2$ is the scaling factor corresponding to the star (where $R$ is the radius of the star and $D$ is
the distance to the star) and $\sigma_{o,i}$ is the error in the observed flux. The SEDs of BSS are constructed using 12 to 20 photometric data points.

We used the Kurucz stellar models \citep{castelli97} to fit the SEDs of BSS. In order to fit the SED to each of the BSS, we kept the $T\mathrm{_{eff}}$ and log $g$ values as free parameters, and chose to range them from 5000--50000 K and 
3--5, respectively. We fixed the metallicity, [Fe/H] = 0.0, the closest option possible in the Kurucz model to the value of metallicity reported of the cluster in \citet{jacobson11}, i.e. [Fe/H] = +0.02$\pm$0.04. The SED fit tool of VOSA returns five best-fits with minimum $\chi^2_r$ values. We closely examined these five best-fit SEDs to see if there was excess in the UV and/or in the IR data-points. If there was excess detected in the IR data-points, we examined the 2MASS images using $Aladin$\footnote{https://aladin.u-strasbg.fr/} to check if there were nearby sources within 2--3$\arcsec$ which would affect the flux values. In BSS1, we found the presence of multiple nearby sources. Hence, we do not present its SED in this work. In other sources with IR excess, we first fitted the optical and IR parts of the SEDs ignoring the UV data points. For sources showing residual flux greater than 30\% in the UV wavelengths in more than one UV data-points, we used a python code, Binary\_SED\_Fitting\footnote{https://github.com/jikrant3/Binary\_SED\_Fitting}, by \citet{jadhav21} to fit two component SEDs.

\begin{figure*} 
\includegraphics[scale =0.25]{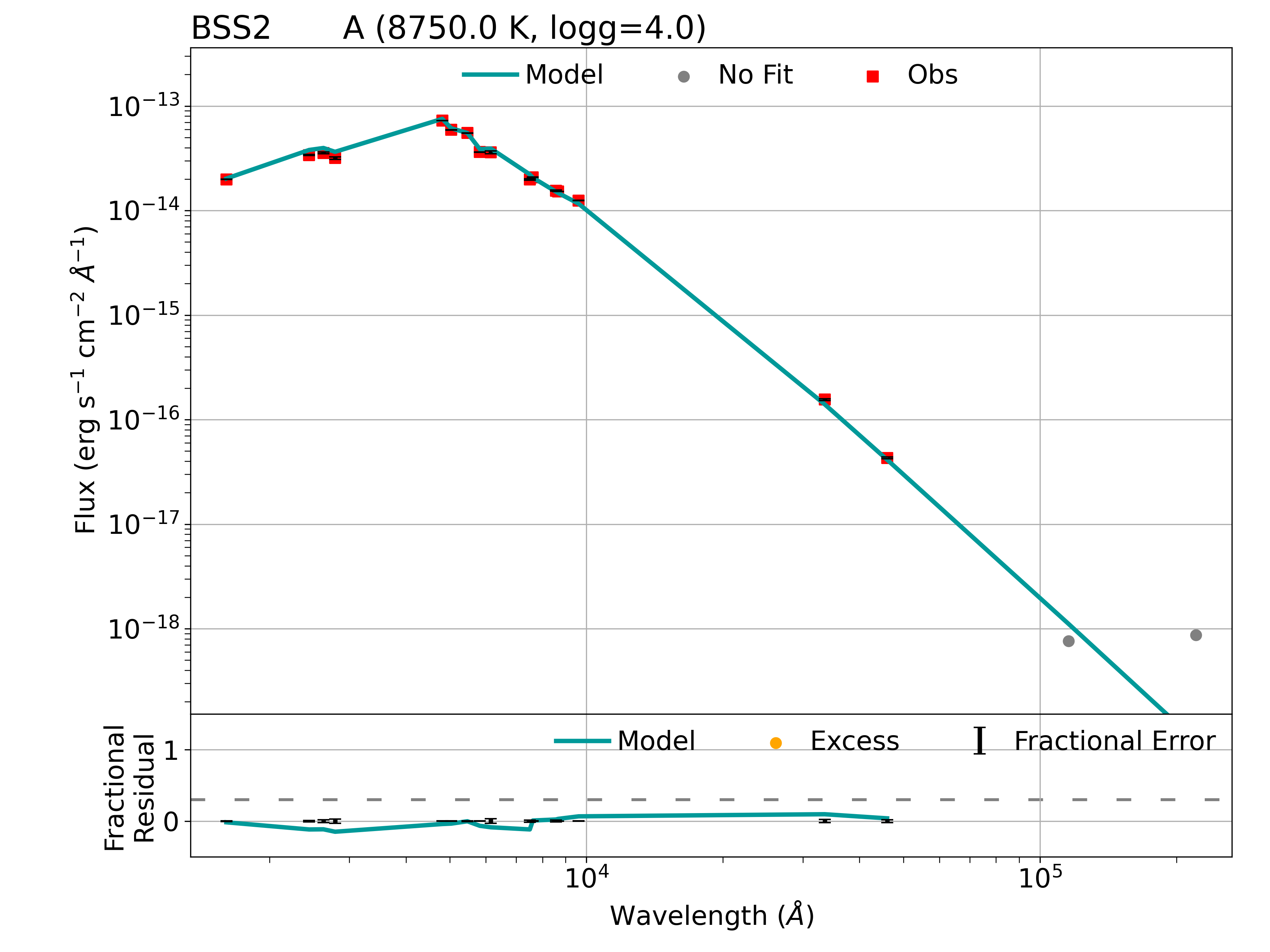} 
\includegraphics[scale =0.25]{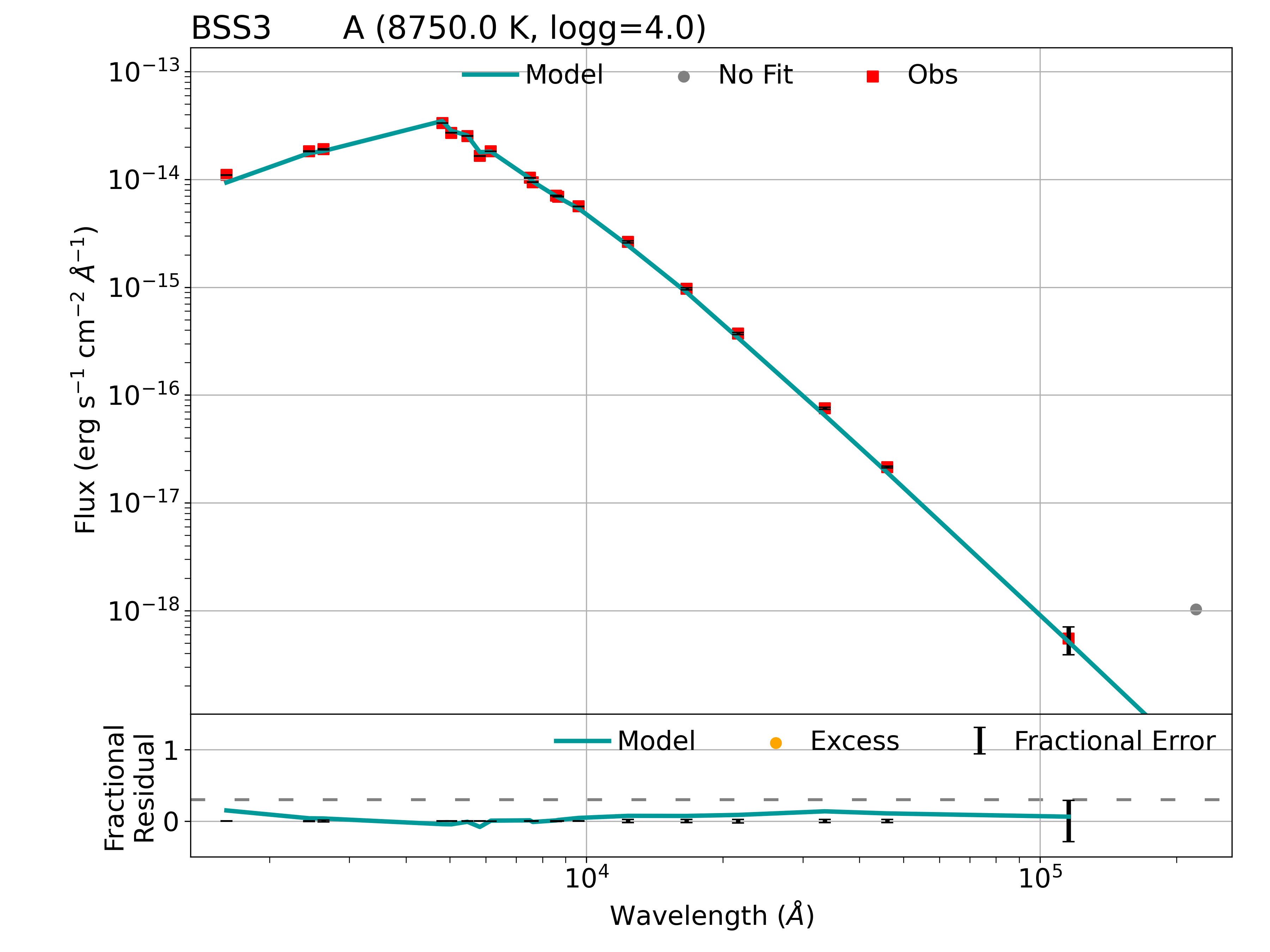} 
\includegraphics[scale =0.25]{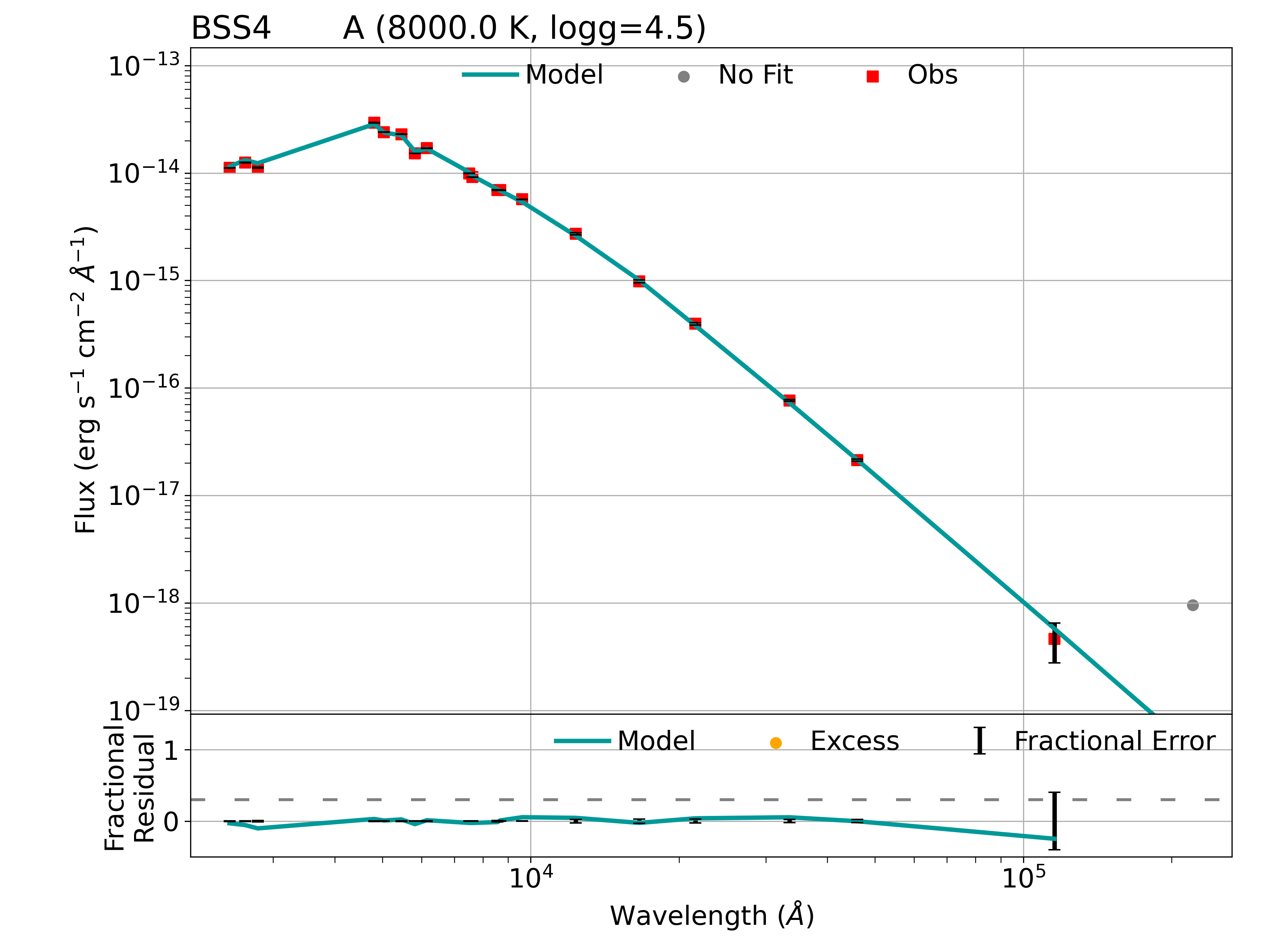} 
\includegraphics[scale =0.25]{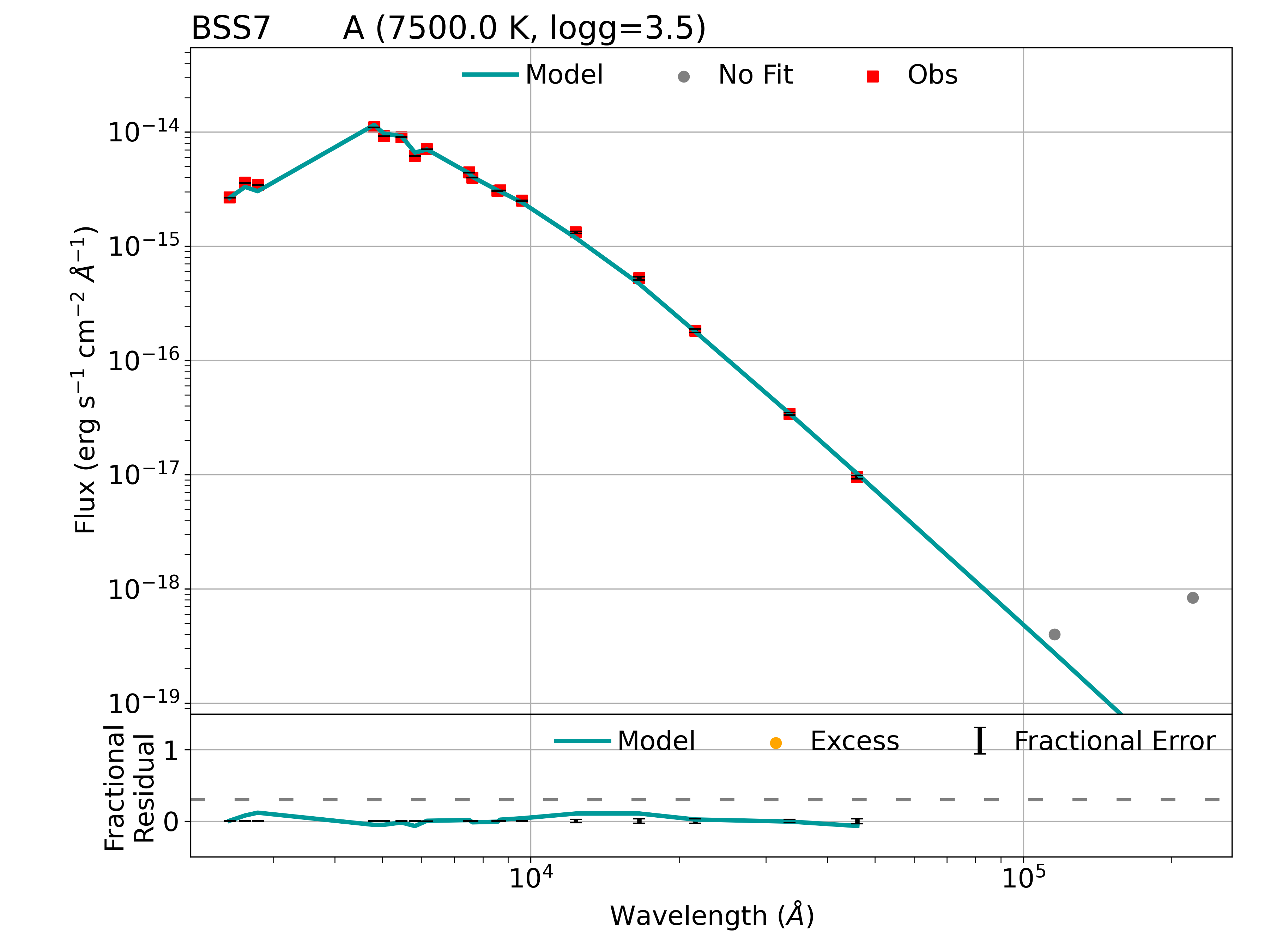} 
\includegraphics[scale =0.25]{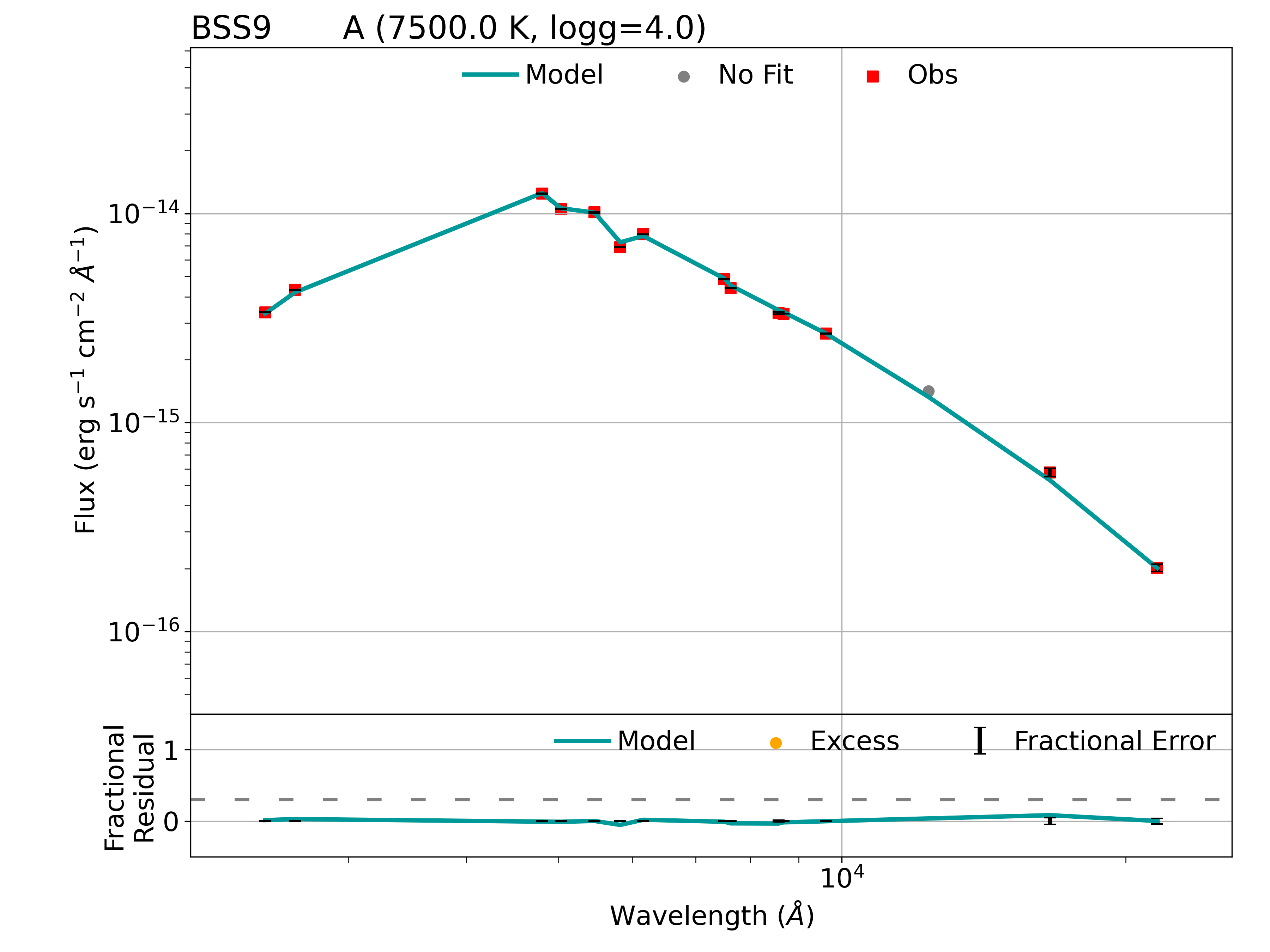} 
\includegraphics[scale =0.25]{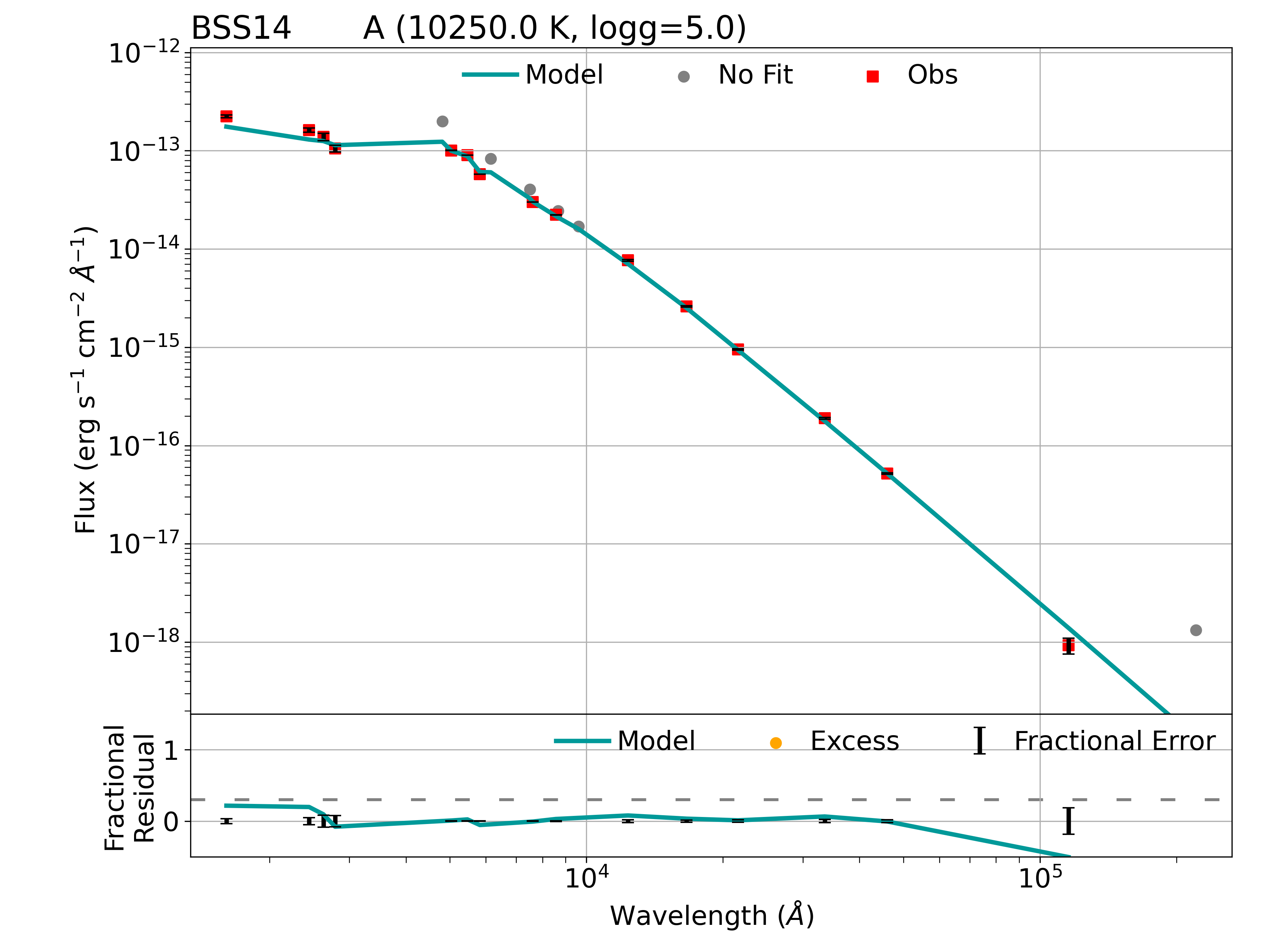}
\includegraphics[scale =0.25]{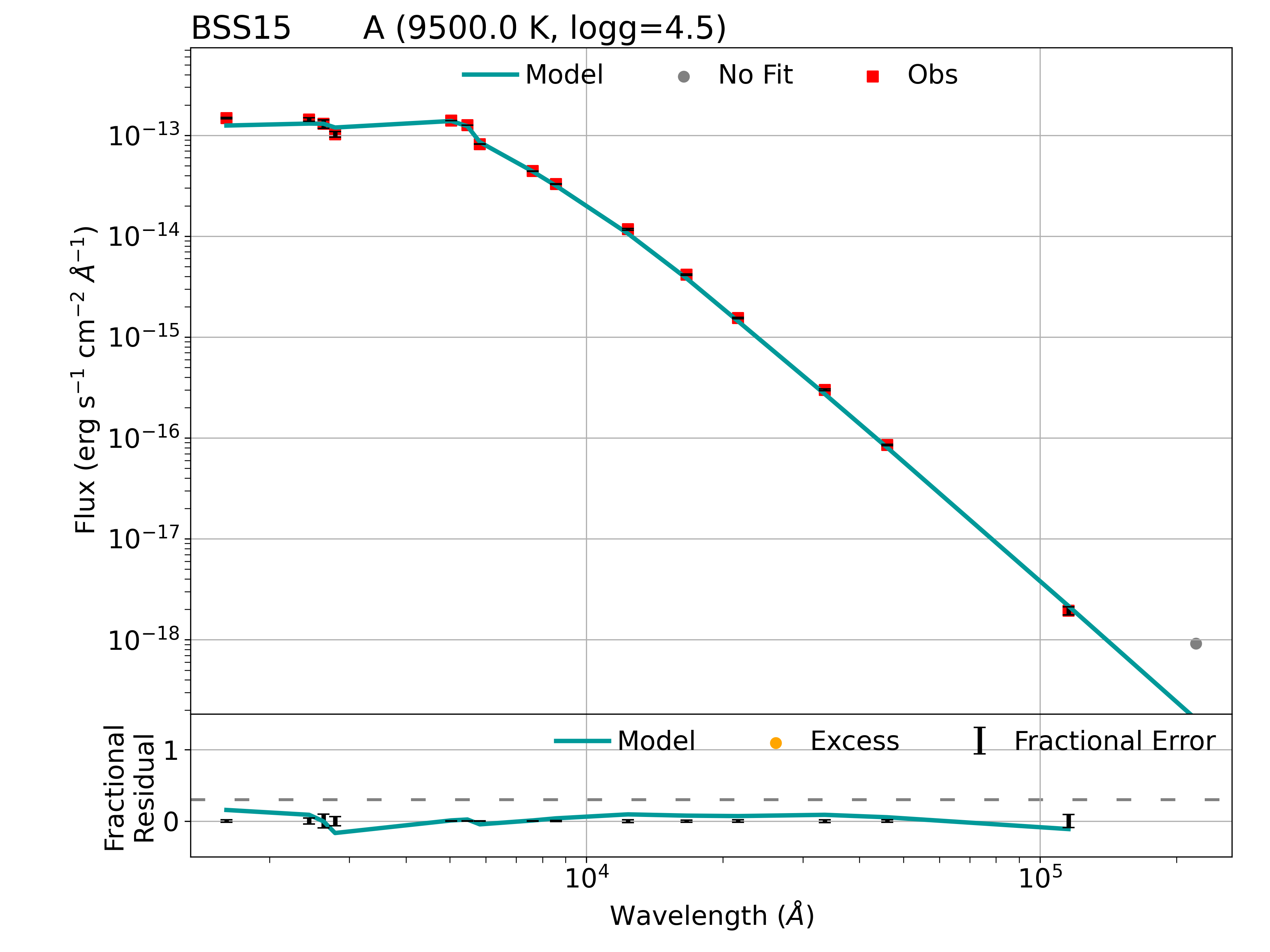} 
\includegraphics[scale =0.25]{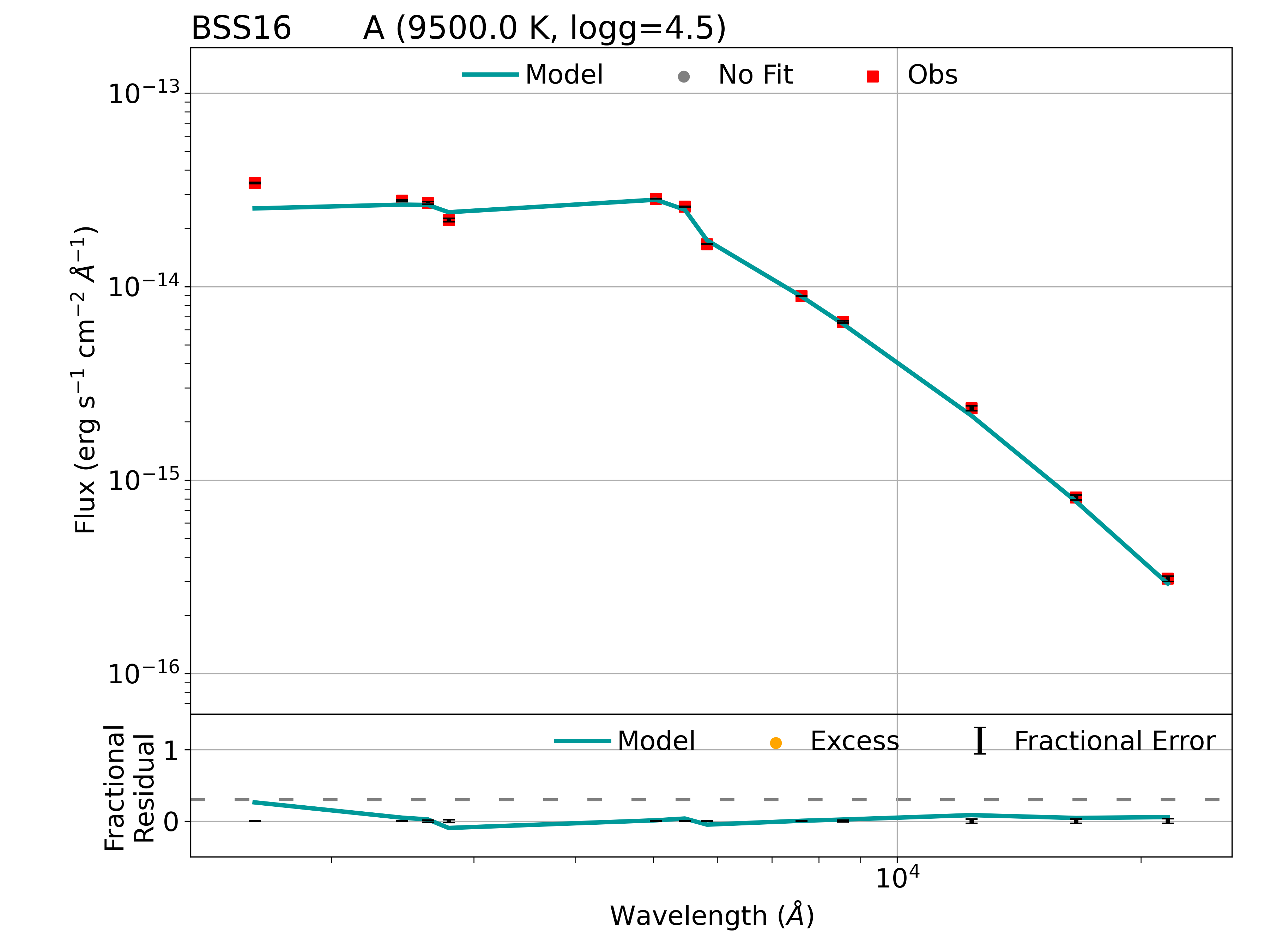} 
\caption{The single-component fit SEDs of BSS. The top panel shows the extinction corrected SEDs fitted with Kurucz stellar models.
The data points used in the fitting are shown as red points with error bars, and data points not used in the fitting are shown as grey points. The bottom panel shows the fractional residual across the filters with a dashed horizontal at a fractional residual of 0.3 to mark the threshold in the excess.}
\label{Fig.2}
\end{figure*}

Out of the 15 BSS candidates, we find that 8 BSS can be successfully fitted with a single component SED. All of these 8 BSS candidates show negligible to zero residual between the fitted model and the extinction corrected observed fluxes across all the filters. The fitted spectra to these 8 BSS candidates are shown in the Figure~\ref{Fig.2}. The top panel in the figure shows the model fitted to the extinction corrected flux values. The data points used in the fitting are shown as red points with error bars, where the error bars represent the errors in the observed fluxes. Since the errors in the observed fluxes are very small for most filters, the error bars are smaller than the data points. The data points not included in the fitting either because they were flagged for poor photometric quality or because only upper limits are available, are shown as grey points. For example, in most of the BSS, only upper limits to fluxes are available in the \textit{WISE} W4 filter, and hence these are excluded from the fits. Also, in BSS10 and BSS14, the PanSTARRS data points are not included in the fitting as they were flagged for poor photometric quality. The bottom panel shows the fractional residual in each filter. A dashed horizontal line at a fractional residual equal to 0.3 demarcates the threshold beyond which the excess is considered for a two-component fit. 

\begin{figure*} 
\includegraphics[scale =0.25]{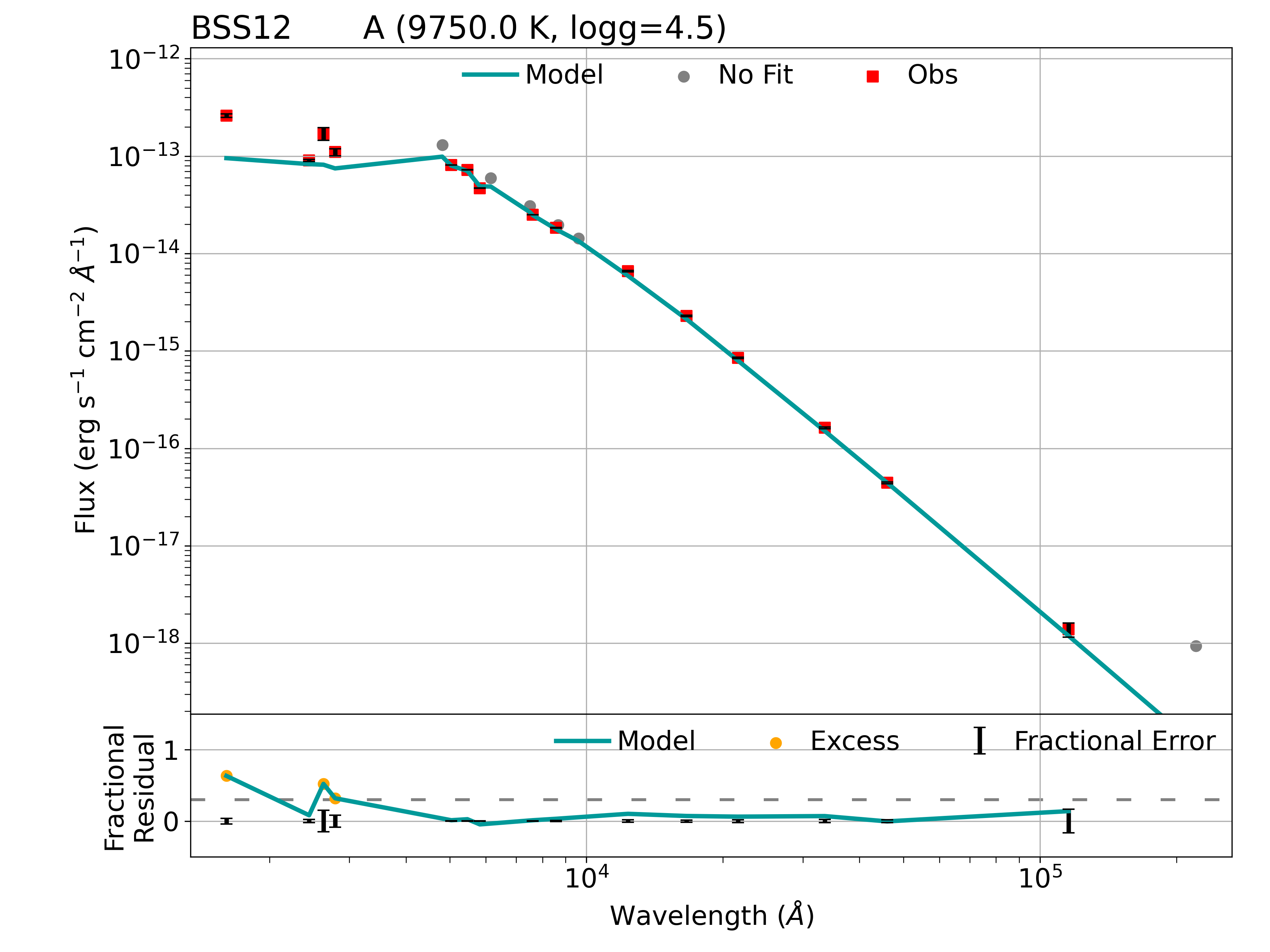} 
\includegraphics[scale =0.25]{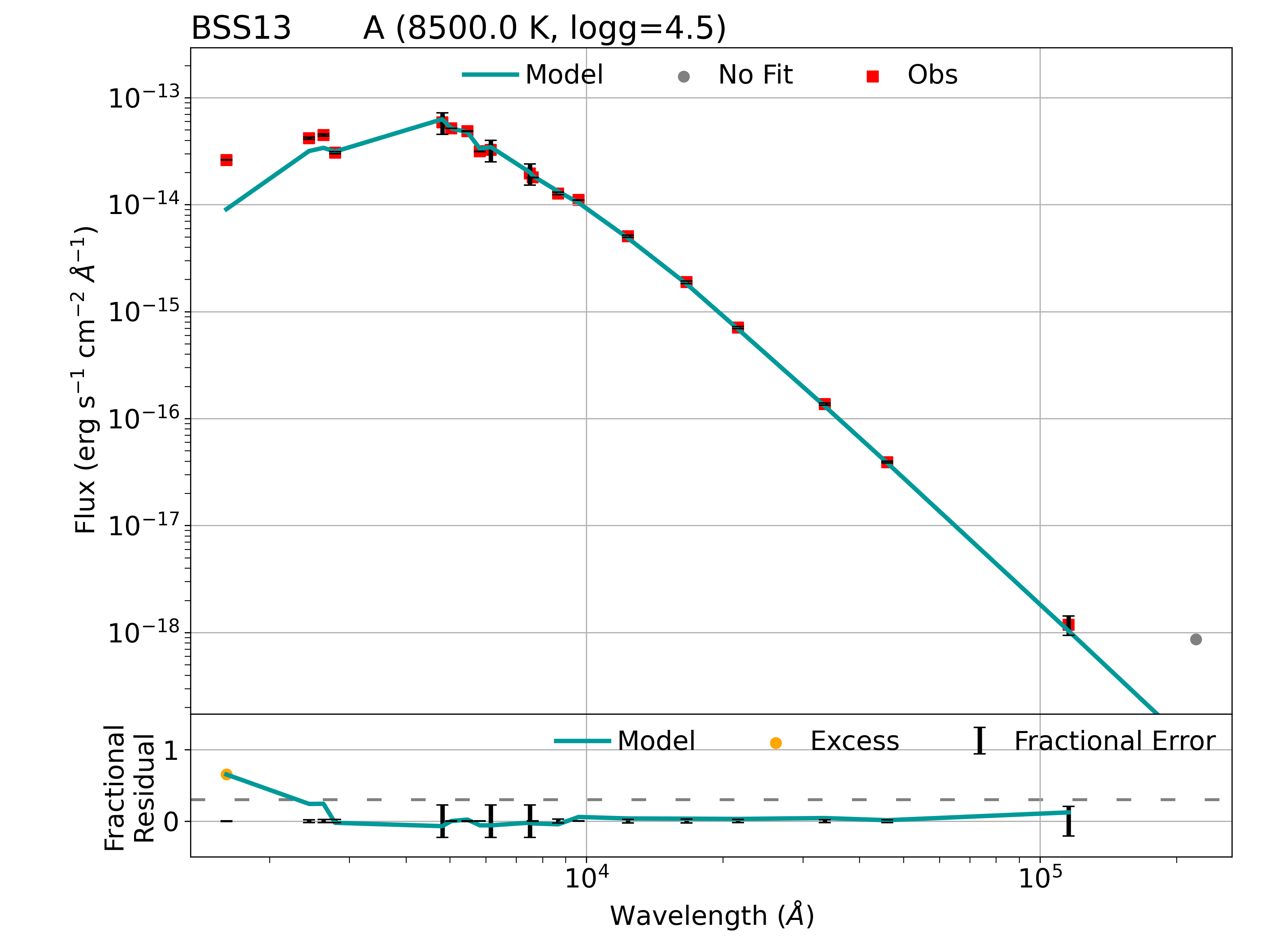}
\caption{The single-component fit SEDs of BSS as in Figure~\ref{Fig.2}. A fractional residual greater than 0.3 is found in the FUV filter F169M in both BSS12 and BSS13.}
\label{Fig.3}
\end{figure*}

As can be seen in Figure~\ref{Fig.2}, the residuals in the fits are close to zero in all these BSS candidates. Despite the fitted model reproducing the observed values quite well, the $\chi^2_r$ values of the SEDs are large (see Table~\ref{table 3}). As the flux errors are very small, even minor deviations in the model from the observed values appear large as compared to the errors, hence by contributing to large $\chi^2_r$. VOSA calculates modified reduced $\chi^2$, \textit{vgf$_{b}$}, to estimate the \textit{visual goodness of fit}. It is calculated by assuming the observational errors in the fluxes ($\sigma_{o,i}$) to be at least 10\% of the observed fluxes ($F_{o,i}$). This parameter is helpful to know how closely the model reproduces the observations and is particularly useful if the photometric errors of any of the catalogues used to build the SED are underestimated. As per VOSA, the SED fits with \textit{vgf$_{b}$} values smaller than 10--15 are good fits. In several previous studies, such as  \citet{2019NatAs...3..553R}, \citet{2019AJ....157...78J}, and \citet{2021MNRAS.506.5201R},
\textit{vgf$_{b} <$} 15 is used as a criterion to select reliable SED fits. We obtain \textit{vgf$_{b}$} values to be less than 1, except for BSS12, BSS13, and BSS14, suggesting that the high reduced $\chi^2$ value are driven by the small errors in the photometric data points. BSS12 and BSS13 have excess in UV data points, as can be seen in Figure~\ref{Fig.3}. Therefore, it is not surprising that their SED fits have somewhat larger \textit{vgf$_{b}$}, $\lesssim$5. We conclude that our SED fits, and the derived fundamental parameters of the BSS candidates, listed in Table~\ref{table 3}, are reliable. Since the fractional residual exceeds 0.3 in BSS12 and BSS13 in one or two UV filters, these two BSS were considered for two-component fits. However, when we attempted to fit them two-component SEDs as detailed in Section \ref{Section 3.4}, we were unable to obtain reliable fits for them as models having a large range of temperatures fitted the observed SEDs.  These two BSS need to be further investigated for characterizing their properties and to understand the reason for their UV excess.

\subsubsection{Two-component SED fits} \label{Section 3.4}

Of 7 BSS showing fractional residual greater than 0.3 in the UV data-points, we successfully fitted 5 BSS with two-components. Of these five BSS, BSS5 and BSS10 are known SB1 \citep{nine20}, whereas BSS6, BSS8, and BSS11 are classified as single members due to an absence of detectable periodic radial velocity variation \citep{nine20}. BSS6 is also a rapid rotator according to \citet{nine20} as well as a known $\delta$ Scuti variable \citep{mochejska99}. For stars without radial velocity variation, the UV flux can be from a hot companion in a low inclination orbit. For genuine single members, the excess in UV may be due to hot spots, chromospheric, or coronal activities. The presence of such activities will be implied if the BSS are detected in X-rays. We checked for X-ray counterparts of these 5 BSS showing UV excess. None of these sources are detected as X-ray sources by XMM-Newton. This indicates that the UV excess is very likely to be due to a hotter companion. 

To fit the second component in these 5 BSS, we used the Koester WD model \citep{koester10} which allows a temperature range of 5000--80000 K and log $g$ range of 6.5--9.5. 
The parameters of the cool components including their errors are derived from the VOSA fits, whereas the parameters of the hot component are derived from the composite fit. In order to estimate the errors in the parameters of the hot components, we used a statistical approach as followed by \citet{jadhav21}. We added Gaussian noise proportional to the errors to each data point, and generated 100 iterations of observed SEDs for each BSS. These 100 SEDs were fitted with double components. Then the median values of the parameters derived from the 100 SED fits were considered as the parameters of the hot companions, and the standard deviation from the median parameters were considered as errors in the parameters.
When temperature of all the 100 fits converge to one value or the standard deviation is within $\pm$250 K, we assume the error in the temperature to be 250 K which is half of the temperature step size of the Kurucz models. Table~\ref{table 3} lists the parameters of the hot components along with these estimated errors as well as the parameters of the associated BSS.
The double component SED fits for the BSS are shown in the Figure~\ref{Fig.4}. The top panel for each BSS shows the fitted SED, with the cool component in orange dashed line, hot component in blue dashed line, the residuals of 100 iterations are shown in light pink lines, and the composite fit is shown in green solid line. The bottom panel shows the residual for both single and composite fits. It can be seen that by fitting the second component the residual in the UV data-point significantly decreases (see Table~\ref{table 3}). 

{\it BSS5 (WOCS 20009) -} BSS5 is a long-period SB1 with an orbital period of 4190 days \citep{nine20}. The best-fit single SED had the $\chi^2_r$ of 500 which reduced to a considerably lower value of 101 (\textit{vgf$_{b}$} = 0.09) with the two-component fit SED. The best fit parameters of the hot companion, T$\mathrm{_{eff}}$ = 15000 K, L = 0.258 L$\mathrm{_{\odot}}$, and R = 0.075 R$\mathrm{_{\odot}}$, suggest that it could be a low-mass WD.  

{\it BSS6 (WOCS 25008) -} BSS6 was classified as a single member by \citet{nine20} that has rapid rotation. It is also a known $\delta$ Scuti variable with a pulsation period of 0.0868 days \citep{mochejska99}. The two-component SED fitting of this BSS reduced the $\chi^2_r$ of the single temperature fit from 396 to 67 (\textit{vgf$_{b}$} = 0.59). The hot companion has an estimated temperature of T$\mathrm{_{eff}}$ = 11750 K. The estimated L= 0.672 L$\mathrm{_{\odot}}$ and R = 0.197 R$\mathrm{_{\odot}}$ of the hot companion, are slightly larger as compared to that of a WD.

{\it BSS8 (WOCS 27010) -} BSS8 has no radial velocity variation \citep{nine20}. The two-component SED fit significantly reduced the $\chi^2$ of the fit, from 879 of the single temperature SED, to 79 (\textit{vgf$_{b}$} = 0.087). The best fit parameters of the hot companion, T$\mathrm{_{eff}}$ = 15500 K, L = 0.253 L$\mathrm{_{\odot}}$, and R = 0.069 R$\mathrm{_{\odot}}$, suggest that it could be a low mass (LM) WD.  

{\it BSS10 (WOCS 10011) -} BSS10 is an SB1 with an orbital period equal to 517 days and an eccentricity of 0.73 \citep{nine20}. The two-component SED fit significantly reduced the $\chi^2$ of the fit, from 722 of the single temperature SED, to 143 (\textit{vgf$_{b}$} = 0.53). The best fit parameters of the hot companion are T$\mathrm{_{eff}}$ = 12500 K, L = 1.293 L$\mathrm{_{\odot}}$, and R = 0.242 R$\mathrm{_{\odot}}$. The L and R of the hot companion are slightly larger as compared to that of a WD.

{\it BSS11 (WOCS 15015) -} BSS11 is a single member according to \citet{nine20}. The two-component SED fit slightly reduced the $\chi^2_r$ of the fit, from 134 of the single temperature SED, to 110 (\textit{vgf$_{b}$} = 0.089). The best fit parameters of the hot companion are T$\mathrm{_{eff}}$ = 15250 K, L = 1.55 L$\mathrm{_{\odot}}$, and R = 0.178 R$\mathrm{_{\odot}}$. The L and R of the hot companion are slightly larger as compared to that of a WD.

\begin{figure*} 
\includegraphics[scale=0.32]{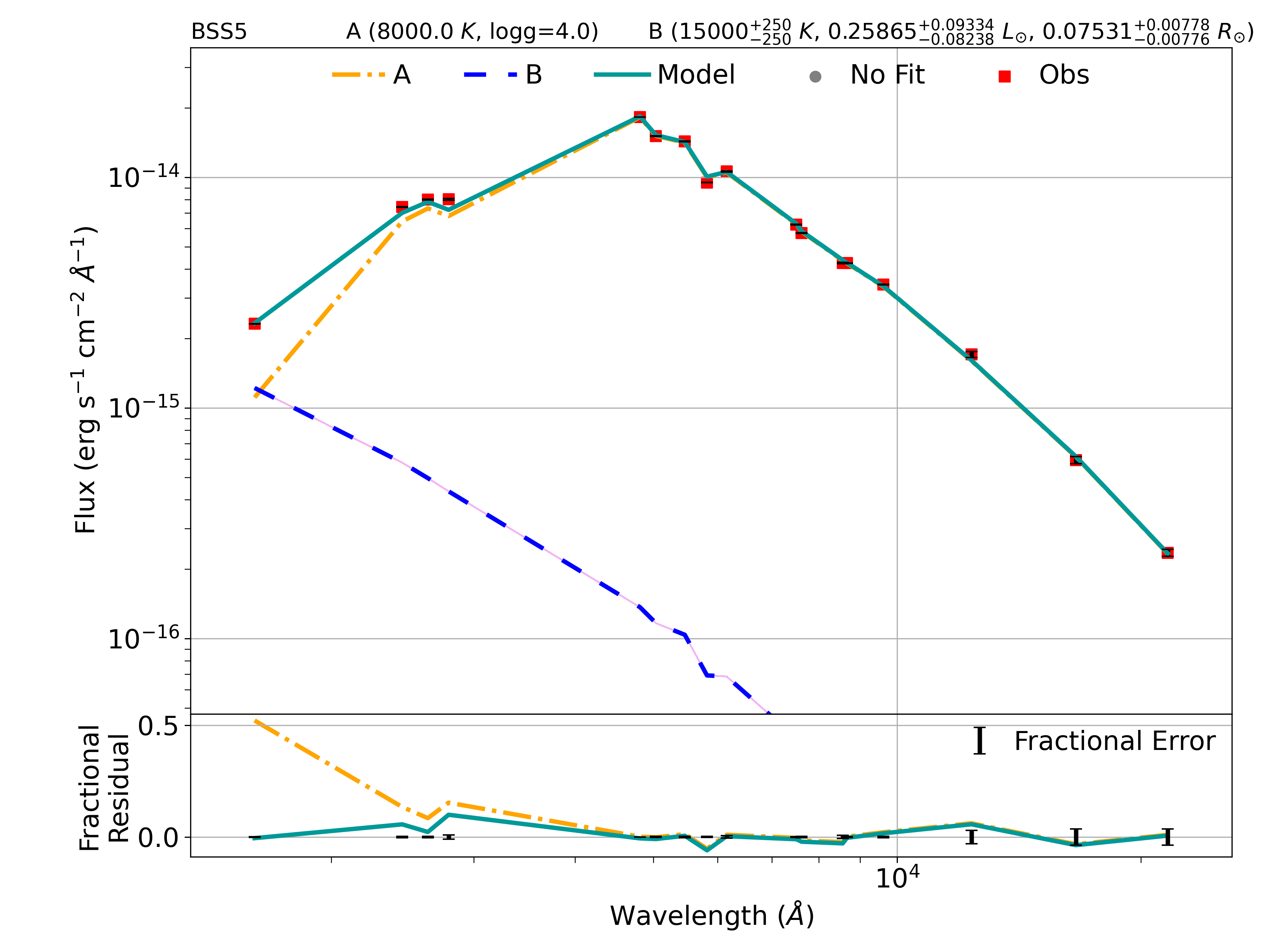} 
\includegraphics[scale=0.32]{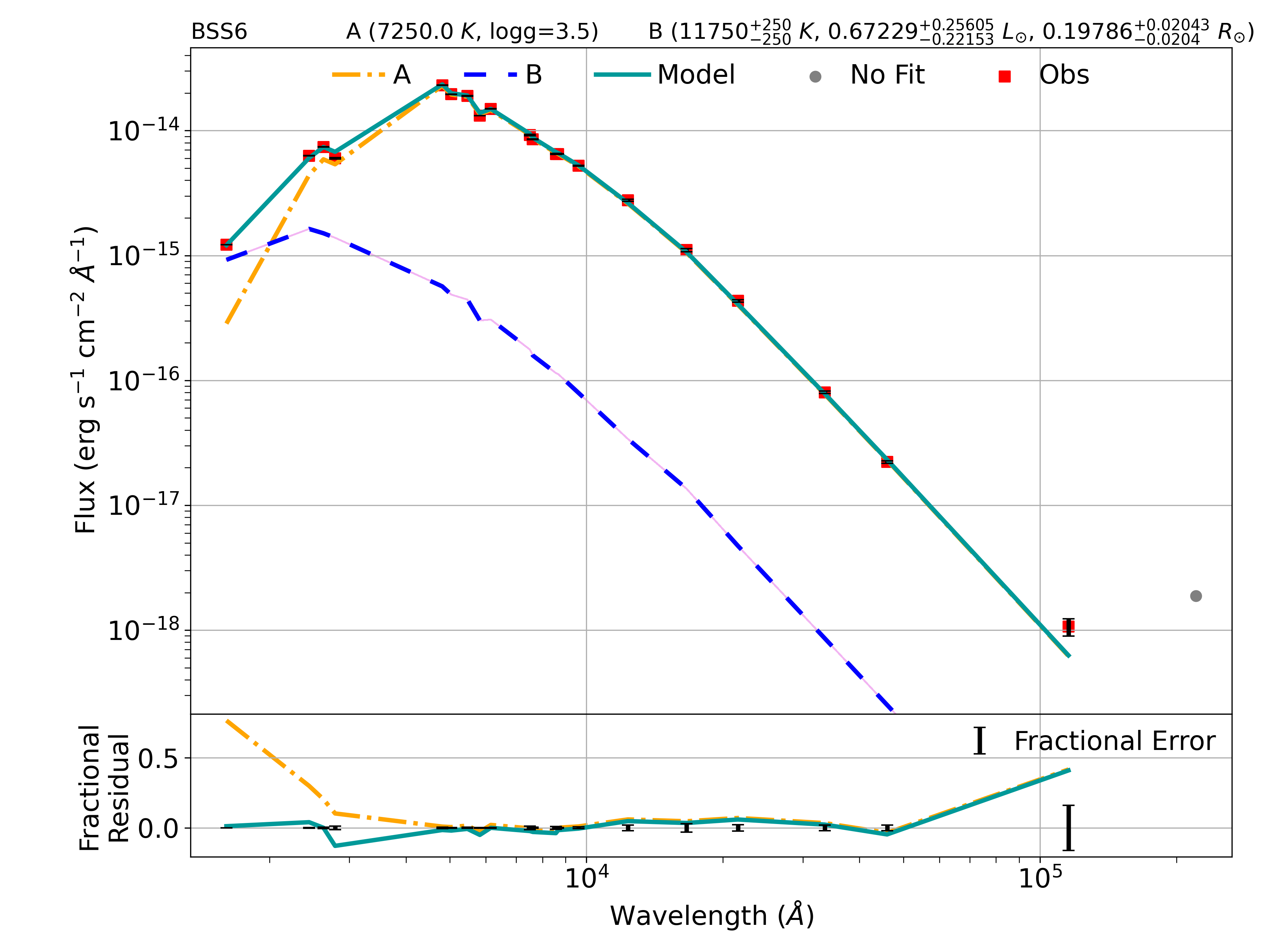} 
\includegraphics[scale=0.32]{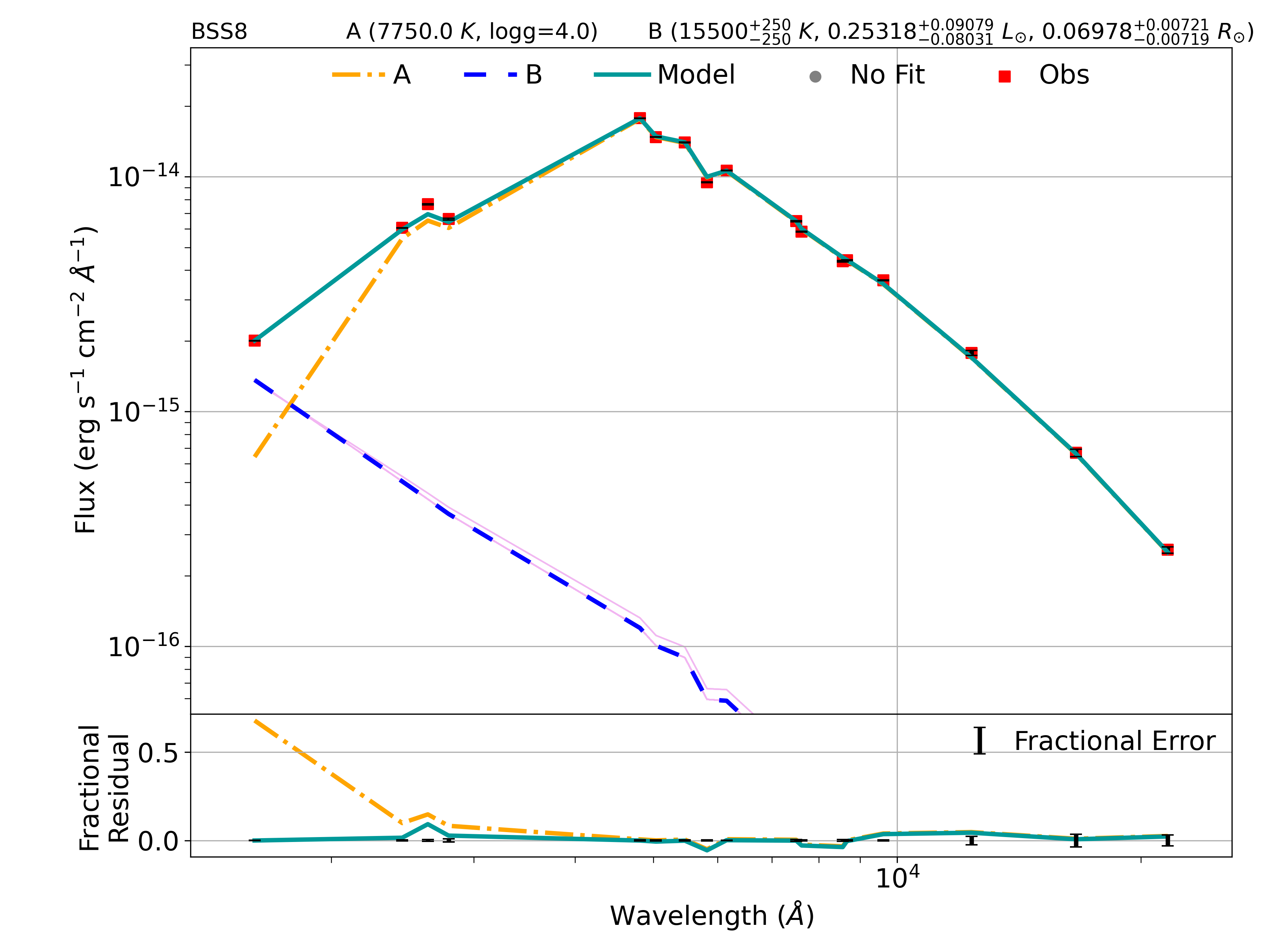}

\caption{Two-component SEDs of BSS. The top panel shows the extinction corrected fluxes and the model SED with the cool (A) component in the orange dashed line, hot (B) component in the blue dashed line along with residuals of iterations shown as light pink lines, and the composite fit in the green solid line. The fitted data points are shown as red points with error bars according to flux errors and the data points not included in the fit are shown as grey data points. The bottom panel shows the fractional residual for both single fit (orange) and composite fit (green). The fractional errors are shown on the x-axis. The parameters of the cool and hot components derived from SED fits, along with their estimated errors are mentioned at the top of the figures.}
 \label{Fig.4}
\end{figure*}

\begin{figure*}  \ContinuedFloat
\includegraphics[scale=0.32]{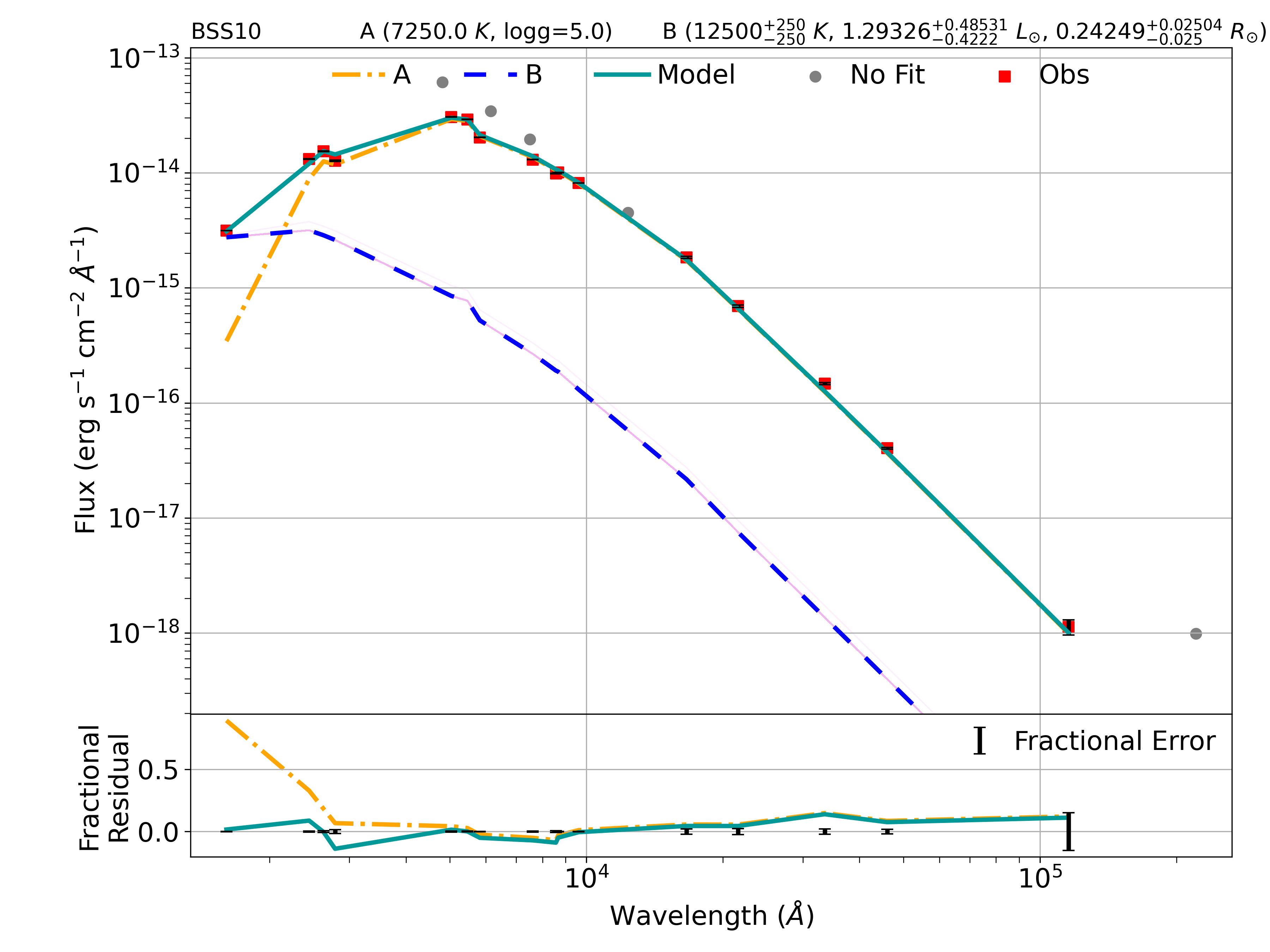} 
\includegraphics[scale=0.32]{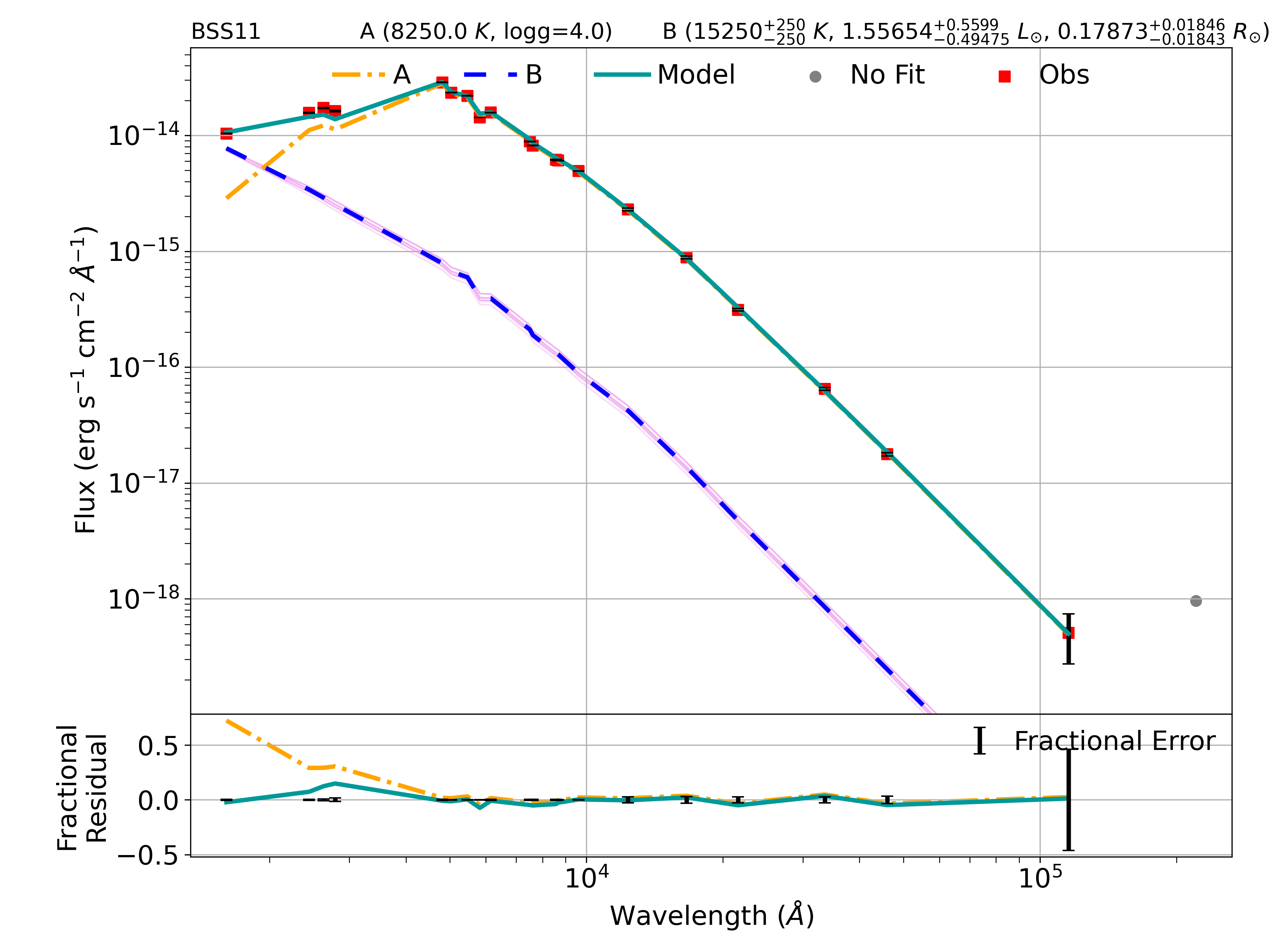}
\caption{The two-component fit SEDs of BSS (cont.)}
\label{Fig.4}
\end{figure*}

\section{Discussion} \label{Section 4}

\subsection{BSS Properties}
In 8 of the BSS candidates, a single-temperature SED fits satisfactorily, whereas in 5 BSS (2 SB1, 1 rapidly rotating star which is a known $\delta$ Scuti variable, and 2 single members), a composite of two components is found as a better fit. The temperatures of the BSS range from 7250 K to 10250 K. This is considerably higher than the BSS of older open clusters such as NGC 188 in which they range from 6100 K to 6800 K \citep{gosnell15}, or King 2, in which they range from 5750 to 8000 K \citep{jadhav21}, but are comparable to another intermediate-age open cluster NGC 2682, in which they range from 6250 K to 9000 K \citep{jadhav21}.
We estimated BSS masses by extrapolating the single star main-sequence 1.6 Gyr PARSEC isochrone. The BSS masses range from 1.50 M$_{\odot}$ to 2.65 M$_{\odot}$. Comparing these masses with the turnoff mass, 1.55 M$_{\odot}$, of the cluster, we have 5 BSS (33\%) with masses greater than 0.5 M$_{\odot}$ from the turnoff mass, 6 BSS (35\%) with masses greater than 0.2--0.5  M$_{\odot}$ from the turnoff mass, 4 BSS (26\%) with masses greater than 0--0.2 M$_{\odot}$ from the turnoff mass, and a single BSS (6\%) with mass slightly less than the turnoff mass. This mass distribution of the BSS looks similar for other open clusters with ages smaller than 2 Gyr as shown by \citet{leiner21}. In particular, our greater percentage (33\%) of massive BSS with masses greater than 0.5 M$_{\odot}$ from the turnoff mass, is consistent with its younger age, and with the lower overall percentage (23\%) of such BSS in the entire sample of clusters of ages 1--10 Gyr as reported by \citet{leiner21}. 

\begin{figure*} 
\includegraphics[scale=0.6]{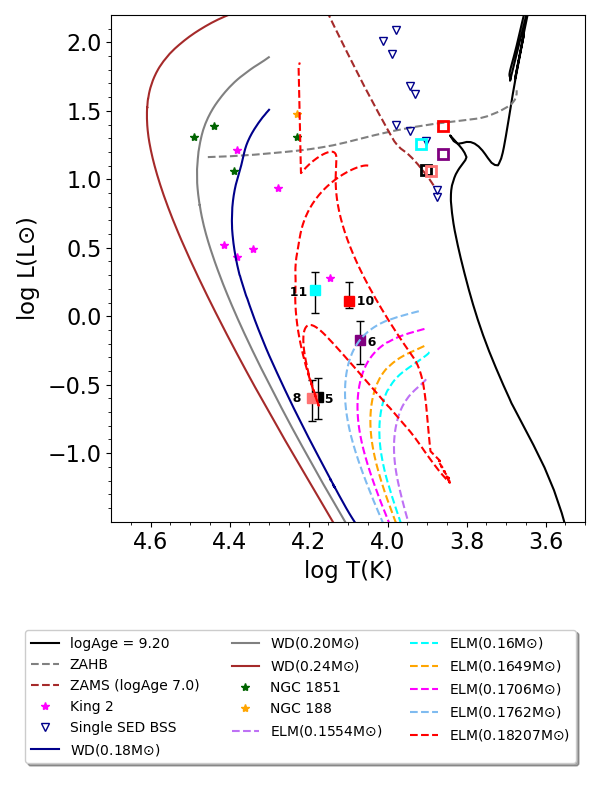} 
\caption{The H-R diagram showing the single component BSS as blue open triangles, the cooler component of two-component BSS as open squares of different colors, and their hotter components marked by filled squares of the same colors as well as numbered according to Table~\ref{table 2}. The isochrone of the cluster age 1.6 Gyr is plotted as black curve, the zero-age main-sequence is shown as brown dashed curve. The zero-age horizontal branch is shown as grey dashed curve. The low-mass WD models of masses 0.18--0.24 M$_{\odot}$ from \citet{panei07} are plotted as solid curves, and the ELM WD models of masses 0.15--0.18 M$_{\odot}$ from \citet{althaus13} are plotted as dashed curves. The hot companions of BSS of King 2 \citep{jadhav21}, NGC 1851 \citep{singh20}, and NGC 188 \citep{subramaniam16} are also shown for reference.}
\label{Fig.5}
\end{figure*}

In order to know the nature of the hot companions, we plot the H-R diagram as shown in the Figure~\ref{Fig.5}.
The PARSEC isochrone of log age = 9.20 is over-plotted on the diagram \citep{bressan12} along with the low-mass WD cooling curves from \citet{panei07}, evolutionary sequences for ELM WD with masses smaller than 0.18 M$_{\odot}$ from \citet{althaus13}, and BaSTI zero age HB \citep[][ZAHB]{hidalgo18}. The BSS with single temperature SEDs are marked as blue open triangles. The cooler components of the BSS fitted with two temperature SEDs are shown as open squares of different colors and their hot companions are shown by the same colored filled squares. As a reference, the hot companions of BSS of other open clusters such as King 2 \citep{jadhav21}, NGC 1851 \citep{singh20}, and NGC 188 \citep{subramaniam16} are shown in the figure. BSS5-A and BSS8-A are lying close to the main-sequence. Their hot companions, BSS5-B and BSS8-B lie on the ELM WD model of 0.1820 M$_{\odot}$ from \citet{althaus13}. Both these companions are thus likely ELM WD candidates. BSS6-A is lying near the blue hook, whereas its hot companion, BSS6-B is lying on the ELM WD track of mass 0.1762 M$_{\odot}$ \citep{althaus13}. It may be a very young ELM He-WD. BSS10-A is lying near the blue hook and BSS11-A is lying away from the main-sequence. Both of their hot companions BSS10-B and BSS11-B are located between the ELM WD tracks of masses 0.1762 and 0.1820 M$_{\odot}$. These hot companions of BSS10 and BSS11 are also likely to be ELM WD stars. 

\subsection{Formation Pathways of BSS with Hot Companions}

The discovery of ELM WD with masses $\leq$ 0.18 M$_{\odot}$ in these five BSS as secondary companions confirm that these objects are indeed post mass transfer systems as He-WD $\leq$ 0.4 M$_{\odot}$ cannot form within the Hubble time from a single star evolution \citep{brown10}. The presence of ELM WD as companions in these BSS imply that their formation happened via the Case-A/Case-B mass transfer process. This formation mechanism would result into a binary system having a short orbital period. The two known SB1 among these BSS with ELM WD companions, BSS5 and BSS10, are long-period binaries. BSS5 is a known SB1 from \citet{nine20} with an orbital period of 4190 days and an eccentricity of 0.27. BSS10 is an SB1 with an orbital period equal to 517 days and an eccentricity of 0.37. These long orbital periods along with the estimated masses of ELM WD imply that these systems are inconsistent with the stable mass transfer process \citep{rappaport95}. We infer that these two BSS are likely in hierarchical triple systems, with BSS progenitor as a member of inner binary which gained mass from its primary companion when it filled its Roche lobe \citep[Kozai mechanism;][]{perets09}. BSS5 has an estimated mass of $\sim$1.65 M$_{\odot}$, whereas BSS10 has an estimated mass of $\sim$1.93 M$_{\odot}$. Assuming that they had an initial mass same as the turnoff mass, $\sim$1.55 M$_{\odot}$, these two BSS gained at least $\sim$0.1--0.4 M$_{\odot}$ from the primary of the triple system which then evolved to become ELM WD. Several examples of LM/ELM WD companions to long orbital period SB1 have been found in the literature. For example, \citet{pandey21} found 3 BSS+WD systems with LM/ELM WD with orbital parameters outside the limits of the stable mass transfer by Case-B process. They also speculated that such systems may in fact be triple systems with the detected LM/ELM WD in a closer inner orbit yet to be characterized. Alternately, the presence of LM/ELM WD in these long orbital period BSS may imply that the stable MT mechanism needs modification to include their formation. 

BSS6 is a known $\delta$ Scuti variable and a rapid rotator \citep{nine20}. Its rapid rotation points toward a recent mass transfer episode. We estimate its mass to be 1.76 M$_{\odot}$. Thus, it has gained an additional mass of at least $\sim$0.2 M$_{\odot}$ in the mass transfer process. Of the five BSS with ELM WD companions, BSS8 and BSS11 do not have radial velocity variation as per \citet{nine20}. We estimated their masses to be 1.65 M$_{\odot}$ and 1.76 M$_{\odot}$, respectively. Hence these two BSS gained additional mass of at least $\sim$0.1--0.2 M$_{\odot}$. Using the grid of ELM WD available in \citet{althaus13}, we estimate the cooling ages of these five ELM WD to range between $\sim$100--1000 Myr. Moreover, we suggest that as all these BSS formed via Case-A/Case-B mass transfer process, they are not likely to have differences in their surface chemical composition since they did not accrete material from an AGB companion.

There are two BSS, BSS12 and BSS13, which show fractional residual greater than 0.3 in the FUV filter. The fitting of the second component in these two BSS using the Koester model resulted into a large range of temperature of the hotter components. Further study is needed to explain the UV excess in these two BSS. Finally, out of the four known SB1s of the cluster, BSS2 and BSS9, are successfully fitted with a single temperature fit. These binary systems may be older mass transfer systems containing cool WDs as companions that are undetected with the current UV observations. 

\section{Conclusions} \label{Section 5}
\begin{enumerate}
\item We studied 16 BSS of the intermediate-age open cluster NGC 7789 using the UVIT observations in three NUV filters N245M, N263M, and N279N, and one FUV filter F148W. We presented a detailed analysis of all the BSS on the basis of their SEDs, excluding of BSS1, which has a few nearby sources within 2$\arcsec$.

\item There are 8 BSS that do not show excess in the UV data points and are successfully fitted with a single temperature SED. Two BSS, BSS12 and BSS13, show some excess in the UV data points, however, we are unable to obtain a good fit for the second component in our current analysis. There are 5 BSS that show UV excess greater than 30\% from the best-fit model and are successfully fitted with a hot+cool composite models. 

\item Based on the SED fitting, temperatures of BSS range from 7250 K to 10250 K. By extrapolating based on the single-star cluster isochrone, BSS masses range from 1.50 M$_{\odot}$ to 2.65 M$_{\odot}$, with 33\% of the BSS having masses greater than 0.5 M$_{\odot}$ from the turnoff mass. 

\item We discover ELM WD candidates with temperatures ranging from T$\mathrm{_{eff}}$ $\sim$11750 K to $\sim$15500 K in five BSS. These ELM WD are likely with masses smaller than $\sim$0.18 M$_{\odot}$. The presence of ELM WD as secondary companions confirm that these are indeed post mass transfer systems. The estimated cooling ages of these ELM WD range from $\sim$100-1000 Myr.

\item Among the five BSS with ELM WD companions, two BSS are known SB1. BSS5 has an orbital period $\geq$ 4000 days. BSS10 has an orbital period of $\geq$500 days. The presence of ELM WD companions demands a close binary orbit for the Case-A/Case-B mass transfer to happen. These BSS are likely to be in the inner binary of triple systems with the detected ELM WD in a closer inner orbit.   

\item The other three BSS with ELM WD companions are having temperatures T$\mathrm{_{eff}}$ $\sim$11750 K--15500 K. One of them, BSS6, is a rapid rotator and a known $\delta$ Scuti variable. 

\item From our analysis, we suggest that at least 33\% BSS formed through mass transfer process in this cluster.
\end{enumerate}

\section*{Acknowledgements}
KV and AP acknowledge the financial support from ISRO under the $AstroSat$ archival data utilization program (No. DS\_2B-13013(2)/3/2019-Sec.2). AS acknowledges the support from SERB grant SPF/2020/000009. This publication uses the data from the $AstroSat$ mission of the Indian Space Research Organisation (ISRO), archived at the Indian Space Science Data Centre (ISSDC). The UVIT is built in collaboration
between IIA, Inter-University Centre for Astronomy and Astrophysics (IUCAA), Tata Institute of Fundamental Research (TIFR), ISRO and Canadian Space Agency (CSA). This work has made use of data from the European Space Agency (ESA) mission Gaia DR2 , processed by the Gaia Data Processing and Analysis Consortium (DPAC 3). This research has made use of the VizieR catalogue access tool, CDS, Strasbourg, France. This research made use of {\small ASTROPY}, a {\small PYTHON} package for astronomy \citep{Astropy2013}, {\small NUMPY} \citep{Harris2020}, {\small MATPLOTLIB} \citep{Hunter4160265}, and {\small SCIPY} \citep{Virtanen2020}. This research also
made use of NASA\'s Astrophysics Data System (ADS 4).

\section*{Data availability}
The data underlying this article are publicly available at \url{https://astrobrowse.issdc.gov.in/astro\_archive/archive/Home.jsp}. The derived data generated in this research will be shared on reasonable request to the corresponding author.

\begin{landscape}
\begin{table}
	\centering
	\caption{The details of UVIT/$AstroSat$ observations, source detections, and counterparts found in members identified by \citet{rao21} and \citet{nine20}.}
	\begin{tabular}{cccccc}
		\hline
		\\
	    Filter & Exposure & PSF & Detections & Counterparts & Counterparts  \\
	    ~ & sec & $\arcsec$ & ~ & \citet{rao21} & \citet{nine20} \\
		\hline
		\\
F169M & 609 & 1.52 & 260 & 84 & 78 \\
N245M & 3219 & 1.08 & 2631  & 1056 & 497 \\ 
N263M & 1142 & 1.34 & 5264 & 1310 & 540 \\
N279N & 1021 & 1.09 & 1712 & 888 & 464 \\

	\\
	
		\hline
		\label{table 1}
	\end{tabular}
\end{table}
\end{landscape}

\begin{landscape}
\begin{table}
	\centering
	\caption{For each BSS, WOCS ID from \citet{nine20} in column 2, coordinates in columns 3 and 4, Gaia EDR3 ID in column 5, whether the BSS is a confirmed single member (SM), binary member (BM), or a candidate (C) in column 6, UVIT F169M flux in column 7, GALEX NUV flux in column 8, UVIT N245M, N63M, and N279N fluxes in columns 9, 10, and 11, respectively. All values of fluxes are in the unit of ergs $\mathrm{s^{-1} cm^{-2} \AA^{-1}}$.}
	\begin{tabular}{llccccccccc}
		\hline
		\\
	    Name & WOCS ID & RA & DEC & Gaia EDR3 source\_id & Member & UVIT.F169M$\pm$err & GALEX.NUV$\pm$err & UVIT.N245M$\pm$err & UVIT.N263M$\pm$err & UVIT.N279N$\pm$err \\
		\hline
		\\
BSS1 & 5004 & 359.37662 & +56.74175 & 1995014924638265472 & SM & 2.78e-15$\pm$1.77e-17 & 5.23e-15$\pm$1.39e-16 & 5.42e-15$\pm$6.86e-17 & 7.24e-15$\pm$1.42e-16 & 7.05e-15$\pm$1.92e-16  \\
BSS2 & 5011 & 359.24595 & +56.79255 & 1995061344646874368 & BM, RR & 2.57e-15$\pm$1.58e-17 & 3.70e-15$\pm$1.12e-16 & 4.84e-15$\pm$5.29e-17 & 6.58e-15$\pm$1.22e-16 & 6.65e-15$\pm$1.90e-16 \\
BSS3 & 10010 & 359.22600 & +56.67969 & 1995012897413751936 & SM & 1.42e-15$\pm$4.98e-18 & 2.22e-15$\pm$9.03e-17 & 2.60e-15$\pm$1.51e-17 & 3.52e-15$\pm$4.03e-17 & -- \\
BSS4 & 16020 & 359.57416 & +56.82777 & 1995021178110590464 & SM & -- & 1.44e-15$\pm$8.39e-17 & 1.58e-15$\pm$6.14e-18 & 2.31e-15$\pm$1.41e-17 & 2.34e-15$\pm$2.30e-17 \\ 	
BSS5 & 20009 & 359.45741 & +56.74361 & 1995014237443527680 & BM &  2.99e-16$\pm$4.83e-19 & 1.36e-15$\pm$8.03e-17 & 1.06e-15$\pm$3.3e-18 & 1.48e-15$\pm$5.1e-18 & 1.68e-15$\pm$1.58e-17 \\
BSS6 & 25008 & 359.37891 & +56.66363 & 1995010801470149504 & SM, RR & 1.57e-16$\pm$1.5e-19 & 7.49e-16$\pm$5.79e-17 & 8.97e-16$\pm$2.25e-18 & 1.36e-15$\pm$6.14e-18 & 1.25e-15$\pm$1.56e-17 \\  	
BSS7 & 25024 & 359.69041 & +56.74441 & 1995016436466871296 & SM, RR & -- & 2.82e-16$\pm$4.13e-17 & 3.79e-16$\pm$5.47e-17 & 6.63e-16$\pm$1.26e-18 & 7.20e-16$\pm$3.35e-17 \\ 	
BSS8 & 27010 & 359.48308 & +56.71208 & 1995010973268848512 & SM & 2.58e-16$\pm$3.91e-19  & -- & 8.63e-16$\pm$2.16e-18 & 1.41e-16$\pm$7.45e-18 & 1.38e-15$\pm$1.14e-17 \\
BSS9 & 36011 & 359.28679 & +56.64375 & 1995010354793197696 & BM & -- &   8.47e-16$\pm$1.04e-16 & 4.80e-16$\pm$8.15e-19 & 7.98e-16$\pm$2.65e-18 & -- \\
BSS10 & 10011 & 359.17554 & +56.73677 & 1995060313854772736 & BM & 4.06e-16$\pm$7.55e-19 & -- & 1.87e-15$\pm$8.66e-18 & 2.84e-15$\pm$2.15e-17 & 2.65e-15$\pm$4.03e-17 \\
BSS11 & 15015 & 359.52192 & +56.78412 & 1995020211733162112 & SM & 1.34e-15$\pm$5.67e-18 & 2.19e-15$\pm$9.23e-17 & 2.23e-15$\pm$1.24e-17 & 3.19e-15$\pm$2.98e-17 & 3.40e-15$\pm$5.46e-17 \\
BSS12 & 4009 & 359.27670 & +56.78061 & 1995061997481928832 & C &  3.35e-14$\pm$1.32e-15 & 1.39e-14$\pm$2.12e-16 & 1.28e-14$\pm$2.88e-16 & 3.15e-14$\pm$7.73e-15 & 2.30e-14$\pm$1.93e-15 \\
BSS13 & 9027 & 359.71512 & +56.65205 & 1995003242327470592 & C, RR &  3.37e-15$\pm$2.05e-17 & 3.71e-15$\pm$1.18e-16  & 5.95e-15$\pm$1.21e-16 & 8.27e-15$\pm$1.85e-16 & 6.41e-15$\pm$1.37e-16  \\
BSS14 & 3009 & 359.23699 & +56.68662 & 1995012897413744384 & SM, VRR &  2.89e-14$\pm$9.81e-16 & 1.50e-14$\pm$2.22e-16 & 2.30e-14$\pm$1.16e-15 & 2.55e-14$\pm$2.17e-15 & 2.21e-14$\pm$1.72e-15 \\
BSS15 & 1014 & 359.20484 & +56.63882 & 1995011832262285568 & C, RR &  1.91e-14$\pm$3.47e-16 & -- & 2.04e-14$\pm$8.44e-16 & 2.40e-14$\pm$2.33e-15 & 2.14e-14$\pm$1.36e-15  \\
BSS16 & 11011 & 359.36180 & +56.80704 & 1995062100562319360 & C, VRR &  4.43e-15$\pm$3.61e-17 & 3.43e-15$\pm$1.13e-16 & 3.96e-15$\pm$2.46e-17 & 5.00e-15$\pm$7.23e-17 & 4.63e-15$\pm$9.68e-17 \\
\\

	\\
	
		\hline
		\label{table 2}
	\end{tabular}
\end{table}
\end{landscape}

\onecolumn
\begin{table}
	\centering
	\caption{The SED fitting parameters of BSS and their hot companions derived from the best-fit SEDs. For each BSS, whether the single component or two component SED is satisfactory in Column 2, log $g$ in Column 3, temperature, radius, and luminosity in Columns 4--6, the scaling factor by which the model has to be multiplied to fit the data in Column 7, the number of data-points fitted in Column 8, the reduced $\chi^2_r$ in Column 9, and the modified reduced $\chi^2_r$, \textit{vgf$_{b}$} in Column 10. In case of two-component fits the $\chi^2_r$ values of the single fits are given in the brackets, whereas \textit{vgf$_{b}$} values of only the two-component fits are listed.}
	\begin{tabular}{llllllllll}
		\hline
		\\
	    Name & Component & log $g$ & T$\mathrm{_{eff}}$ & R & L & Scaling Factor & N$_{fit}$ & $\chi^2_r$ & \textit{vgf$_{b}$} \\
	    ~ & ~ & ~ & [K] & [R$_{\odot}$] & [L$_{\odot}$] & ~ & ~ & ($\chi^2_{r,single}$) \\
		\hline
		\\
BSS2 & single & 4.0 & 8750$\pm$125 & 3.04$\pm$0.152 & 47.7$\pm$4.81 & 1.18E-21 & 18 & 286.1 & 0.77  \\
BSS3 & single & 4.0 & 8750$\pm$125 & 2.09$\pm$0.104 & 22.4$\pm$2.25 & 5.46E-22 & 20 & 289.5 & 0.51   \\
BSS4 & single & 4.5 & 8000$\pm$125 & 2.28$\pm$0.114 & 19.0$\pm$1.91 & 6.61E-22 & 20 & 86.7 & 0.19 \\ BSS5 & A & 4 & 8000$\pm$125 & 1.79$\pm$0.089 & 11.75$\pm$1.18 & 4.082E-22 & 13 & 101.0 (500.8) & 0.091   \\
~ & B & 9.5 & 15000$_{-250}^{+250}$ & 0.075$\pm$0.007 & 0.258$_{-0.082}^{+0.093}$ & 7.213E-25 & -- & -- \\ 
BSS6 & A & 3.5 & 7250$\pm$125 & 2.50$\pm$0.125 & 15.32$\pm$1.54 & 7.82E-22 & 17 & 67.9 (396.7) & 0.59 \\
~ & B & 9.5 & 11750$_{-250}^{+250}$ & 0.197$\pm$0.020 & 0.672$_{-0.221}^{+0.255}$ & 4.98E-24 & -- & -- \\ 
BSS7 & single & 3.5 & 7500$\pm$125 & 1.62$\pm$0.081 & 7.43$\pm$0.74 & 3.86E-22 & 20 & 278.5 & 0.38 \\ 
BSS8 & A & 4 & 7750$\pm$125 & 1.90$\pm$0.094 & 11.52$\pm$1.16 & 4.55E-22 & 13 & 79.2 (879) & 0.087  \\
~ & B & 9 & 15500$_{-250}^{+250}$ & 0.069$\pm$0.007 & 0.253$_{-0.080}^{+0.090}$ & 6.193E-25 & -- & -- \\ 
BSS9 & single & 4 & 7500$\pm$125 & 1.72$\pm$0.086 & 8.37$\pm$0.84 & 3.77E-22 & 14 & 79.7 & 0.12 \\
BSS10 & A & 5 & 7250$\pm$125 & 3.13$\pm$0.156 & 24.4$\pm$2.45 & 1.25E-21 & 13 & 143.7 (722.4) & 0.53  \\
~ & B & 9.5 & 12500$_{-250}^{+250}$ & 0.242$\pm$0.025 & 1.293$_{-0.42}^{+0.48}$ & 7.47E-24 & -- & -- \\ 
BSS11 & A & 4 & 8250$\pm$125 & 2.08$\pm$0.104 & 18.0$\pm$1.82 & 5.51E-22 & 17 & 110.3 (134.5) & 0.089 \\
~ & B & 9.5 & 15250$_{-250}^{+250}$ & 0.178$\pm$0.018 & 1.556$_{-0.49}^{+0.55}$ & 4.06E-24 & -- & -- \\ 
BSS12 & single & 4.5 & 9750$\pm$125 & 2.99$\pm$0.149 & 81.6$\pm$8.31 & 1.14E-21 & 16 & 64.1 & 5.12  \\
BSS13 & single & 4.5 & 8500$\pm$125 & 2.97$\pm$0.148 & 42.1$\pm$4.57 & 1.13E-21 & 20 & 747.5 & 3.32  \\
BSS14 & single & 5 & 10250$\pm$125 & 3.16$\pm$0.158 & 102.1$\pm$10.3 & 1.28E-21 & 16 & 59.1 & 1.47  \\
BSS15 & single & 4.5 & 9500$\pm$125 & 4.08$\pm$0.204 & 122.8$\pm$12.3 & 2.12E-21 & 16 & 47.4 & 0.82  \\
BSS16 & single & 4.5 & 9500$\pm$125 & 1.84$\pm$0.092 & 25.0$\pm$2.50 & 4.29E-22 & 12 & 158.3 & 1.08 \\
 \\

		\hline
		\label{table 3}
	\end{tabular}
\end{table}
\twocolumn


\newpage








\begin{thebibliography}{99}

\bibitem[\protect\citeauthoryear{Althaus, Miller Bertolami, \& C{\'o}rsico}{2013}]{althaus13} Althaus L.~G., Miller Bertolami M.~M., C{\'o}rsico A.~H., 2013, A\&A, 557, A19. doi:10.1051/0004-6361/201321868
\bibitem[\protect\citeauthoryear{Astropy Collaboration et al.}{2013}]{Astropy2013} Astropy Collaboration, Robitaille T.~P., Tollerud E.~J., Greenfield P., Droettboom M., Bray E., Aldcroft T., et al., 2013, A\&A, 558, A33. doi:10.1051/0004-6361/201322068
\bibitem[\protect\citeauthoryear{Bayo et al.}{2008}]{bayo08} Bayo A., Rodrigo C., Barrado Y Navascu{\'e}s D., Solano E., Guti{\'e}rrez R., Morales-Calder{\'o}n M., Allard F., 2008, A\&A, 492, 277. doi:10.1051/0004-6361:200810395
\bibitem[\protect\citeauthoryear{Boffin et al.}{2015}]{boffin15} Boffin, H. M. J., Carraro, G., \& Beccari, G., 2015, Astrophysics and Space Science Library, 413. doi:10.1007/978-3-662-44434-4
\bibitem[\protect\citeauthoryear{Bressan et al.}{2012}]{bressan12} Bressan A., Marigo P., Girardi L., Salasnich B., Dal Cero C., Rubele S., Nanni A., 2012, MNRAS, 427, 127. doi:10.1111/j.1365-2966.2012.21948.x
\bibitem[\protect\citeauthoryear{Brown et al.}{2010}]{brown10} Brown W.~R., Kilic M., Allende Prieto C., Kenyon S.~J., 2010, ApJ, 723, 1072. doi:10.1088/0004-637X/723/2/1072
\bibitem[\protect\citeauthoryear{Burbidge \& Sandage}{1958}]{burbidge58} Burbidge E.~M., Sandage A., 1958, ApJ, 128, 174
\bibitem[\protect\citeauthoryear{Cantat-Gaudin et al.}{2018}]{cantat18} Cantat-Gaudin T., Jordi C., Vallenari A., Bragaglia A., Balaguer-N{\'u}{\~n}ez L., Soubiran C., Bossini D., et al., 2018, A\&A, 618, A93. doi:10.1051/0004-6361/201833476
\bibitem[\protect\citeauthoryear{Cantat-Gaudin \& Anders}{2020}]{cantat20} Cantat-Gaudin T., Anders F., 2020, A\&A, 633, A99. doi:10.1051/0004-6361/201936691
\bibitem[\protect\citeauthoryear{Castelli, Gratton, \& Kurucz}{1997}]{castelli97} Castelli F., Gratton R.~G., Kurucz R.~L., 1997, A\&A, 318, 841
\bibitem[\protect\citeauthoryear{Chen \& Han}{2008}]{chen08} Chen X., Han Z., 2008, MNRAS, 387, 1416. doi:10.1111/j.1365-2966.2008.13334.x
\bibitem[\protect\citeauthoryear{Chen et al.}{2017}]{chen17} Chen X., Maxted P.~F.~L., Li J., Han Z., 2017, MNRAS, 467, 1874. doi:10.1093/mnras/stx115

\bibitem[\protect\citeauthoryear{Cohen, Wheaton, \& Megeath}{2003}]{cohen03} Cohen M., Wheaton W.~A., Megeath S.~T., 2003, AJ, 126, 1090. doi:10.1086/376474
\bibitem[\protect\citeauthoryear{de Juan Ovelar et al.}{2020}]{overlar20} de Juan Ovelar M., Gossage S., Kamann S., Bastian N., Usher C., Cabrera-Ziri I., Dotter A., et al., 2020, MNRAS, 491, 2129. doi:10.1093/mnras/stz3128
\bibitem[\protect\citeauthoryear{Ferraro, Fusi Pecci, \& Bellazzini}{1995}]{ferraro95} Ferraro F.~R., Fusi Pecci F., Bellazzini M., 1995, A\&A, 294, 80
\bibitem[\protect\citeauthoryear{Fitzpatrick}{1999}]{fitzpatrick99} Fitzpatrick E.~L., 1999, PASP, 111, 63. doi:10.1086/316293

\bibitem[\protect\citeauthoryear{Friel et al.}{1995}]{friel95} Friel E.~D., Janes K.~A., Hong L., Lotz J., Tavarez M., 1995, fmw..conf, 189

\bibitem[\protect\citeauthoryear{Fusi Pecci et al.}{1992}]{fusi92} Fusi Pecci, F., Ferraro,F. R., Corsi,C. E., Cacciari, C. \& Buonanno, R. 1992, AJ, 104, 1831
\bibitem[\protect\citeauthoryear{Gaia Collaboration et al.}{2018}]{brown18} Gaia Collaboration, Brown A.~G.~A., Vallenari A., Prusti T., de Bruijne J.~H.~J., Babusiaux C., Bailer-Jones C.~A.~L., et al., 2018, A\&A, 616, A1. doi:10.1051/0004-6361/201833051
\bibitem[\protect\citeauthoryear{Gaia Collaboration et al.}{2021}]{brown21} Gaia Collaboration, Brown A.~G.~A., Vallenari A., Prusti T., de Bruijne J.~H.~J., Babusiaux C., Biermann M., et al., 2021, A\&A, 649, A1. doi:10.1051/0004-6361/202039657

\bibitem[\protect\citeauthoryear{Gim et al.}{1998}]{gim98} Gim M., Vandenberg D.~A., Stetson P.~B., Hesser J.~E., Zurek D.~R., 1998, PASP, 110, 1318. doi:10.1086/316266
\bibitem[\protect\citeauthoryear{Gim et al.}{1998}]{gim98a} Gim M., Hesser J.~E., McClure R.~D., Stetson P.~B., 1998, PASP, 110, 1172. doi:10.1086/316241
\bibitem[\protect\citeauthoryear{Gosnell et al.}{2014}]{gosnell14} Gosnell N.~M., Mathieu R.~D., Geller A.~M., Sills A., Leigh N., Knigge C., 2014, ApJL, 783, L8. doi:10.1088/2041-8205/783/1/L8
\bibitem[\protect\citeauthoryear{Gosnell et al.}{2015}]{gosnell15} Gosnell N.~M., Mathieu R.~D., Geller A.~M., Sills A., Leigh N., Knigge C., 2015, ApJ, 814, 163. doi:10.1088/0004-637X/814/2/163

\bibitem[\protect\citeauthoryear{Harris et al.}{2020}]{Harris2020} Harris C.~R., Millman K.~J., van der Walt S.~J., Gommers R., Virtanen P., Cournapeau D., Wieser E., et al., 2020, Natur, 585, 357. doi:10.1038/s41586-020-2649-2
\bibitem[\protect\citeauthoryear{Hidalgo et al.}{2018}]{hidalgo18} Hidalgo S.~L., Pietrinferni A., Cassisi S., Salaris M., Mucciarelli A., Savino A., Aparicio A., et al., 2018, ApJ, 856, 125. doi:10.3847/1538-4357/aab158
\bibitem[\protect\citeauthoryear{Hills \& Day}{1976}]{hills76} Hills J.~G., Day C.~A., 1976, ApL, 17, 87
\bibitem[\protect\citeauthoryear{Hunter}{2007}]{Hunter4160265}Hunter J. D., 2007, Computing in Science Engineering, 9, 90
\bibitem[\protect\citeauthoryear{Indebetouw et al.}{2005}]{indebetouw05} Indebetouw R., Mathis J.~S., Babler B.~L., Meade M.~R., Watson C., Whitney B.~A., Wolff M.~J., et al., 2005, ApJ, 619, 931. doi:10.1086/426679

\bibitem[\protect\citeauthoryear{Jadhav, Sindhu, \& Subramaniam}{2019}]{jadhav19} Jadhav V.~V., Sindhu N., Subramaniam A., 2019, ApJ, 886, 13. doi:10.3847/1538-4357/ab4b43
 \bibitem[\protect\citeauthoryear{Jadhav et al.}{2021}]{jadhav21} Jadhav V.~V., Pandey S., Subramaniam A., Sagar R., 2021, JApA, 42, 89. doi:10.1007/s12036-021-09746-y

\bibitem[\protect\citeauthoryear{Jacobson et al.}{2011}]{jacobson11} Jacobson H.~R., Pilachowski C.~A., Friel E.~D., 2011, AJ, 142, 59. doi:10.1088/0004-6256/142/2/59
\bibitem[\protect\citeauthoryear{Jim{\'e}nez-Esteban, Solano, \& Rodrigo}{2019}]{2019AJ....157...78J} Jim{\'e}nez-Esteban F.~M., Solano E., Rodrigo C., 2019, AJ, 157, 78. doi:10.3847/1538-3881/aafacc
\bibitem[\protect\citeauthoryear{Johnson \& Sandage}{1955}]{johnson55} Johnson H.~L., Sandage A.~R., 1955, ApJ, 121, 616
\bibitem[\protect\citeauthoryear{Leigh \& Sills}{2011}]{leigh11} Leigh N., Sills A., 2011, MNRAS, 410, 2370. doi:10.1111/j.1365-2966.2010.17609.x
\bibitem[\protect\citeauthoryear{Koester}{2010}]{koester10} Koester D., 2010, MmSAI, 81, 921
\bibitem[\protect\citeauthoryear{Leiner et al.}{2019}]{leiner19} Leiner E., Mathieu R.~D., Vanderburg A., Gosnell N.~M., Smith J.~C., 2019, ApJ, 881, 47. doi:10.3847/1538-4357/ab2bf8
\bibitem[\protect\citeauthoryear{Leiner \& Geller}{2021}]{leiner21} Leiner E.~M., Geller A., 2021, ApJ, 908, 229. doi:10.3847/1538-4357/abd7e9
\bibitem[\protect\citeauthoryear{Leonard}{1989}]{leonard89} Leonard P.~J.~T., 1989, AJ, 98, 217. doi:10.1086/115138

\bibitem[\protect\citeauthoryear{Mapelli et al.}{2009}]{mapelli09}
Mapelli, M., Ripamonti, E., Battaglia, G., Tolstoy, E., Irwin, M. J., Moore, B. \&  Sigurdsson, S. 2009, MNRAS, 396, 1771
\bibitem[\protect\citeauthoryear{Martin et al.}{2005}]{martin05} Martin D.~C., Fanson J., Schiminovich D., Morrissey P., Friedman P.~G., Barlow T.~A., Conrow T., et al., 2005, ApJL, 619, L1. doi:10.1086/426387

\bibitem[\protect\citeauthoryear{McCrea}{1964}]{mccrea64} McCrea W.~H., 1964, MNRAS, 128, 147. doi:10.1093/mnras/128.2.147
\bibitem[\protect\citeauthoryear{McNamara \& Solomon}{1981}]{mcnamara81} McNamara B.~J., Solomon S., 1981, A\&AS, 43, 337
\bibitem[\protect\citeauthoryear{Mochejska \& Kaluzny}{1999}]{mochejska99} Mochejska B.~J., Kaluzny J., 1999, AcA, 49, 351
\bibitem[\protect\citeauthoryear{Momany et al.}{2007}]{momany07}
Momany, Y., Held, E. V., Saviane, I., Zaggia, S., Rizzi, L. \&  Gullieuszik, M. 2007, A\&A, 468, 973
\bibitem[\protect\citeauthoryear{Nine et al.}{2020}]{nine20} Nine A.~C., Milliman K.~E., Mathieu R.~D., Geller A.~M., Leiner E.~M., Platais I., Tofflemire B.~M., 2020, yCat, J/AJ/160/169
\bibitem[\protect\citeauthoryear{Overbeek et al.}{2015}]{overbeek15} Overbeek J.~C., Friel E.~D., Jacobson H.~R., Johnson C.~I., Pilachowski C.~A., M{\'e}sz{\'a}ros S., 2015, AJ, 149, 15. doi:10.1088/0004-6256/149/1/15
\bibitem[\protect\citeauthoryear{Pandey, Subramaniam, \& Jadhav}{2021}]{pandey21} Pandey S., Subramaniam A., Jadhav V.~V., 2021, MNRAS, 507, 2373. doi:10.1093/mnras/stab2308

\bibitem[\protect\citeauthoryear{Panei et al.}{2007}]{panei07} Panei J.~A., Althaus L.~G., Chen X., Han Z., 2007, MNRAS, 382, 779. doi:10.1111/j.1365-2966.2007.12400.x
\bibitem[\protect\citeauthoryear{Perets \& Fabrycky}{2009}]{perets09} Perets H.~B., Fabrycky D.~C., 2009, ApJ, 697, 1048. doi:10.1088/0004-637X/697/2/1048
\bibitem[\protect\citeauthoryear{Preston \& Snedon}{2000}]{preston00}
Preston, G. W. \& Sneden, C. 2000, AJ, 120, 1014
\bibitem[\protect\citeauthoryear{Postma \& Leahy}{2017}]{postma17} Postma J.~E., Leahy D., 2017, PASP, 129, 115002. doi:10.1088/1538-3873/aa8800
\bibitem[\protect\citeauthoryear{Postma \& Leahy}{2020}]{postma20} Postma J.~E., Leahy D., 2020, PASP, 132, 054503. doi:10.1088/1538-3873/ab7ee8
\bibitem[\protect\citeauthoryear{Rao et al.}{2021}]{rao21} Rao K.~K., Vaidya K., Agarwal M., Bhattacharya S., 2021, MNRAS, 508, 4919. doi:10.1093/mnras/stab2894

\bibitem[\protect\citeauthoryear{Rain, Ahumada, \& Carraro}{2021}]{rain21} Rain M.~J., Ahumada J.~A., Carraro G., 2021, A\&A, 650, A67. doi:10.1051/0004-6361/202040072
\bibitem[\protect\citeauthoryear{Rappaport et al.}{1995}]{rappaport95} Rappaport S., Podsiadlowski P., Joss P.~C., Di Stefano R., Han Z., 1995, MNRAS, 273, 731. doi:10.1093/mnras/273.3.731
\bibitem[\protect\citeauthoryear{Rebassa-Mansergas et al.}{2019}]{2019NatAs...3..553R} Rebassa-Mansergas A., Parsons S.~G., Dhillon V.~S., Ren J., Littlefair S.~P., Marsh T.~R., Torres S., 2019, NatAs, 3, 553. doi:10.1038/s41550-019-0746-7
\bibitem[\protect\citeauthoryear{Rebassa-Mansergas et al.}{2021}]{2021MNRAS.506.5201R} Rebassa-Mansergas A., Solano E., Jim{\'e}nez-Esteban F.~M., Torres S., Rodrigo C., Ferrer-Burjachs A., Calcaferro L.~M., et al., 2021, MNRAS, 506, 5201. doi:10.1093/mnras/stab2039
\bibitem[\protect\citeauthoryear{Sahu et al.}{2019}]{sahu19} Sahu S., Subramaniam A., Simunovic M., Postma J., C{\^o}t{\'e} P., Kameswera Rao N., Geller A.~M., et al., 2019, ApJ, 876, 34. doi:10.3847/1538-4357/ab11d0
\bibitem[\protect\citeauthoryear{Sandage}{1953}]{sandage53} Sandage A.~R., 1953, AJ, 58, 61
\bibitem[\protect\citeauthoryear{Sandage}{1962}]{sandage62} Sandage A., 1962, ApJ, 135, 333
\bibitem[\protect\citeauthoryear{Sarajedini}{1993}]{sarajedini93} Sarajedini A., 1993, ASPC,  14, ASPC...53
\bibitem[\protect\citeauthoryear{Sindhu et al.}{2019}]{sindhu19} Sindhu N., Subramaniam A., Jadhav V.~V., Chatterjee S., Geller A.~M., Knigge C., Leigh N., et al., 2019, ApJ, 882, 43. doi:10.3847/1538-4357/ab31a8
\bibitem[\protect\citeauthoryear{Sindhu et al.}{2020}]{sindhu20} Sindhu N., Subramaniam A., Geller A.~M., Jadhav V., Knigge C., Simunovic M., Leigh N., et al., 2020, IAUS, 351, 482. doi:10.1017/S1743921319006975
\bibitem[\protect\citeauthoryear{Singh et al.}{2020}]{singh20} Singh G., Sahu S., Subramaniam A., Yadav R.~K.~S., 2020, ApJ, 905, 44. doi:10.3847/1538-4357/abc173

\bibitem[\protect\citeauthoryear{Subramaniam et al.}{2016}]{subramaniam16} Subramaniam A., Sindhu N., Tandon S.~N., Kameswara Rao N., Postma J., C{\^o}t{\'e} P., Hutchings J.~B., et al., 2016, ApJL, 833, L27. doi:10.3847/2041-8213/833/2/L27
\bibitem[\protect\citeauthoryear{Stetson}{1987}]{stetson87} Stetson P.~B., 1987, PASP, 99, 191. doi:10.1086/131977
\bibitem[\protect\citeauthoryear{Thogersen et al.}{1993}]{thogersen93} Thogersen E.~N., Friel E.~D., Fallon B.~V., 1993, PASP, 105, 1253. doi:10.1086/133304
\bibitem[\protect\citeauthoryear{Tandon et al.}{2017a}]{tandon17} Tandon S.~N., Subramaniam A., Girish V., Postma J., Sankarasubramanian K., Sriram S., Stalin C.~S., et al., 2017, AJ, 154, 128. doi:10.3847/1538-3881/aa8451
\bibitem[\protect\citeauthoryear{Tandon et al.}{2017b}]{tandon17b} Tandon S.~N., Hutchings J.~B., Ghosh S.~K., Subramaniam A., Koshy G., Girish V., Kamath P.~U., et al., 2017, JApA, 38, 28. doi:10.1007/s12036-017-9445-x
\bibitem[\protect\citeauthoryear{Tandon et al.}{2020}]{tandon20} Tandon S.~N., Postma J., Joseph P., Devaraj A., Subramaniam A., Barve I.~V., George K., et al., 2020, AJ, 159, 158. doi:10.3847/1538-3881/ab72a3
\bibitem[\protect\citeauthoryear{Virtanen et al.}{2020}]{Virtanen2020} Virtanen P., Gommers R., Oliphant T.~E., Haberland M., Reddy T., Cournapeau D., Burovski E., et al., 2020, NatMe, 17, 261. doi:10.1038/s41592-019-0686-2
\bibitem[\protect\citeauthoryear{Wang \& Chen}{2019}]{wang19} Wang S., Chen X., 2019, ApJ, 877, 116. doi:10.3847/1538-4357/ab1c61
\bibitem[\protect\citeauthoryear{Webbink}{1976}]{webbink76} Webbink R.~F., 1976, ApJ, 209, 829. doi:10.1086/154781
\bibitem[\protect\citeauthoryear{Wright et al.}{2010}]{wright10} Wright E.~L., Eisenhardt P.~R.~M., Mainzer A.~K., Ressler M.~E., Cutri R.~M., Jarrett T., Kirkpatrick J.~D., et al., 2010, AJ, 140, 1868. doi:10.1088/0004-6256/140/6/1868


\end{thebibliography}
\end{document}